\documentclass[preprint,aps,shownopacs,preprintnumbers,amsmath,amssymb,nofootinbib,epsf,epsfig]
{revtex4}
\usepackage{amstext,amssymb}
\usepackage{amsmath}
\usepackage{graphicx}
\usepackage[hyperfootnotes=true]{hyperref}
\usepackage{color}
\usepackage{array,multirow}
\usepackage{slashed} % feynman slash notation via \slashed{p}
\usepackage{multirow}
\newcommand{\be}{\begin{equation}}
\newcommand{\ee}{\end{equation}}
\newcommand{\bea}{\begin{eqnarray}}
\newcommand{\eea}{\end{eqnarray}}
\newcommand{\nn}{\nonumber}

\def\s1{\hat s}

\def\U1mt{U(1)_{L_\mu-L_\tau}}

\usepackage{slashed}

\begin{document}
\title{Scalar dark matter, Neutrino mass and Leptogenesis  in a $\rm U(1)_{B-L}$ model}

\author{Subhasmita Mishra$^a$}
\email{subhasmita.mishra92@gmail.com}
%\affiliation{\,Department of Physics, IIT Hyderabad,              Kandi - 502285, India }                            
              
\author{Shivaramakrishna Singirala$^b$}
\email{krishnas542@gmail.com}
%\affiliation{\,Discipline of Physics, Indian Institute of Technology Indore, Khandwa Road, Simrol, Indore - 453 552, India}

\author{Suchismita Sahoo$^c$}
\email{suchismita8792@gmail.com}
\affiliation{$^a$Department of Physics, IIT Hyderabad, Kandi - 502285, India\\   $^b$Discipline of Physics, Indian Institute of Technology Indore, Khandwa Road, Simrol, Indore - 453 552, India\\
$^c$Theoretical Physics Division, Physical Research Laboratory, Ahmedabad-380009, India}

% \author[c]{S. Econd,}

\begin{abstract}
We investigate the phenomenology of singlet scalar dark matter in a simple $\rm U(1)_{B-L}$ gauge extension of standard model, made anomaly free with four exotic fermions. The enriched scalar sector and the new gauge boson $Z^\prime$, associated with $\rm U(1)$ gauge extension, connect the dark sector to the visible sector. We compute relic density, consistent with Planck limit and $Z^\prime$ mediated dark matter-nucleon cross section, compatible with PandaX bound. The mass of $Z^\prime$ and the corresponding gauge coupling are constrained from LEP-II and LHC dilepton searches. We also briefly scrutinize  the tree level  neutrino mass with dimension five operator. Furthermore, resonant leptogenesis phenomena is discussed with TeV scale exotic fermions to produce the observed baryon asymmetry of the Universe. Further, we briefly explain the impact of flavor in leptogenesis and we also project the combined constraints on Yukawa,  consistent with oscillation data and observed baryon asymmetry. Additionally, we restrict the new gauge parameters by using the existing data on branching ratios of rare $B(\tau)$ decay modes. We see that the constraints from dark sector are much more stringent from flavor sector. 
\end{abstract}  

\maketitle

%==============================================
\section{Introduction}
%==================================================

Standard Model (SM) of particle physics has produced a remarkable success in explaining physics of the fundamental particles below electroweak scale. However, it does not accommodate the explanation for existence of dark matter (DM), observed matter-asymmetry and few anomalies associated with $B$-sector. The experimental detection of dark matter signal is one of the most awaited event to happen, ever since it was proposed by Fitz Zwicky in early 1930's \cite{Zwicky:1937zza,PhysRev.43.147}. The theoretical proposal of Weakly Interacting Massive Particle (WIMP) has received decent attention in the recent past, where it can produce the correct relic density by freeze-out mechanism. Numerous beyond SM scenarios were realized with WIMP kind of dark matter, and explored immensely in literature \cite{Bertone:2004pz,Berlin:2014tja}. The interaction of WIMP with SM particles opens the scope of its detection prospects, through the production of DM particles in colliders or the direct scattering with the nucleus.

     Moreover, Baryon Asymmetry of the Universe (BAU) being a mysterious problem, needs to be investigated in detail in the growing astro-particle experiments. With the necessity of  Sakharov's conditions for baryogenesis, leptogenesis is the most preferable way to fit with the current cosmological observation of the baryon asymmetry, $\Omega_B h^2 = 0.0223\pm 0.0002$ \cite{Aghanim:2018eyx}, which corresponds to $Y_B\equiv {\eta_B/s} \approx 0.86 \times 10 ^{-10}$. Generation of lepton asymmetry comes from the CP violating out of equilibrium decay of heavy particle, which later converts to the baryon asymmetry through sphaleron transitions. In general, lepton asymmetry produced by the  decay of right handed neutrinos has been widely studied in the literature \cite{Buchmuller:2004nz,Plumacher:1996kc,Buchmuller:2000as,Giudice:2003jh,Strumia:2006qk,Davidson:2008bu}. But with one flavor approximation, the lower limit on right-handed neutrino mass ($\gtrsim 10^9$ GeV) corresponds to the Ibarra bound, which is quite impossible to have any experimental signature in coming decades. Of the many attempts made in literature, resonant leptogenesis is the simplest and well known way to generate a successful asymmetry, by bringing down the mass scale, also compatible with the current neutrino oscillation data \cite{Pilaftsis:2003gt}.

     On the other hand, the LHCb as well as Belle and BaBar experiments  have reported discrepancy in the angular observables of rare decay modes, induced by the quark level transitions, $b \to s l^+ l^-$  and $b \to c l \bar{\nu}_l$ over the last few years. These  measurements include disagreements  at the level of $\sim 3 \sigma$  in the decay distribution \cite{Aaij:2014pli, Aaij:2015oid} and  $P_5^\prime$ observable of $ B \to K^* \mu^+ \mu^-$ \cite{Aaij:2013qta,Huang:2018rys,Huang:2018nnq,Aaij:2015oid}. The decay rate of $B_s \to \phi \mu^+ \mu^-$ also show $3\sigma$ discrepancy in the high recoil limit \cite{Aaij:2013aln, Aaij:2015esa}. Additionally, the  lepton universality violating ratios, $R_K \equiv  \Gamma(B^+ \to K^+ \,\mu^+\,\mu^-)/\Gamma(B^+ \to K^+\,e^+\,e^-)$ along with $R_{K^*} \equiv \Gamma (B^0 \to K^{*0} \mu^+\mu^-)/\Gamma(B^0 \to K^{*0} e^+ e^-)$ deviates  at $\sim 2.5$ $\sigma$ level \cite{Aaij:2014ora, Aaij:2019wad, Bobeth:2007dw, Aaij:2017vbb, Prim, Capdevila:2017bsm} and the  $R_{D^{(*)}}\equiv \Gamma(B \to D^{(*)}\, \tau \bar{\nu}_l)/\Gamma(B \to D^{(*)}\, l \bar{\nu}_l)$ ($R_{J/\psi}\equiv \Gamma(B \to J/\psi\, \tau \bar{\nu}_l)/\Gamma(B \to J/\psi\, l \bar{\nu}_l)$), where  $l=e,\,\mu$ ratios disagrees with the SM at the level of $\sim 3.08\sigma$ ($1.7\sigma$) \cite{HFLAV, Aaij:2017tyk, Ivanov:2005fd, Wen-Fei:2013uea}.
     
To resolve the above issues in a common theory, the SM needs to be extended with additional symmetries or particles. Among many beyond SM frameworks, $U(1)$ extensions stand in the front row, when it comes to simplicity. They are fruitful in phenomenological perspective, with minimal particle and parameter content. These kind of models also provide new scalar and gauge bosonic type mediator particles, that communicate visible sector to the additional particle spectrum. This article includes a minimal $\rm U(1)_{B-L}$ gauge extension of the SM to address these experimental conflicts in a model dependent framework. To avoid triangle gauge anomalies, these extensions require neutral fermions with appropriate $\rm B-L$ charges. A solution of adding three heavy fermions with appropriate $B-L$ charges has been explored in \cite{Ma:2014qra,Ma:2015raa,Nomura:2017kih,Geng:2017foe,Das:2017deo,Das:2019fee,Mishra:2019oqq,Bandyopadhyay:2018cwu,Nomura:2017jxb,Nomura:2017vzp,Singirala:2017see,Singirala:2017cch}. In the present context, we go for the choice of adding four exotic fermions with fractional $\rm B-L$ charges \cite{Patra:2016ofq,Nanda:2017bmi,Biswas:2018yus}. Apart from the scalar content required to generate Majorana mass terms to all the exotic fermions, an additional scalar singlet with fractional charge helps in generating the lepton asymmetry after ${\rm B-L}$ symmetry breaking and also provides light neutrino masses at tree level. We explore scalar singlet DM with a fractional ${\rm B-L}$ charge, whose stability is ensured by an additional $Z_2$ symmetry. We also scrutinize the rare $B$ decay modes at one-loop level (via $Z^\prime$ boson) in the present framework and further constrained the new gauge parameter.
     
Very few works in literature are devoted to $B-L$ models with a choice of adding four exotic fermions and hardly in the context of accommodating leptogenesis. The current model provides a platform to address visible, dark and flavor sectors simultaneously. The plan of the paper is as follows. In section II, we  describe the model along with the relevant interaction Lagrangian. We discuss the symmetry breaking pattern, particle mass spectrum and tree level neutrino mass in section III. Section IV gives a detail study of dark matter phenomenology in relic density and direct detection perspective and also impose constraints from collider studies. Resonant leptogenesis  with  quasi degenerate right-handed fermions and the solutions to Boltzmann equations are discussed in section V. A note on flavor effects is also included here. Then in section VI, we additionally constrain the new gauge parameters from $B$ and $\tau$  sectors. Summarization of the model is provided in section VII.

\section{The model framework}
$B-L$ models are self-consistent gauge extensions of SM, free from triangle gauge anomalies with the addition of extra fermions. With the SM fermion content, the triangle anomalies for $[U(1)^3_{B-L}]$ and  $[(\rm gravity)^2 \times U(1)_{B-L}]$ give a non-zero value, i.e., 
\begin{equation}
\mathcal{A}^{\rm SM}_1\left[U(1)^3_{B-L}\right] = -3,~ %\nonumber \\
\mathcal{A}^{\rm SM}_2\left[\mbox{(gravity)}^2 \times U(1)_{B-L}\right] = -3.
\end{equation}
The conventional way is to add three right-handed neutrinos with $B-L$ charges $-1$ for each. Other possible solution is to add three fermions with exotic $B-L$ charges as $-4,-4$ and $+5$ \cite{Ma:2014qra,Ma:2015raa}. It is also possible to cancel the gauge anomalies by adding four additional fermions carrying fractional $B-L$ charges $N_{1R} (-1/3)$,   $N_{2R}(-2/3)$,   $N_{3R}(-2/3)$ and   $N_{4R}(-4/3)$ \cite{Patra:2016ofq}, deriving explicitly below 
\begin{eqnarray}
\mathcal{A}_1\left[U(1)^3_{B-L} \right]&&= \mathcal{A}^{\rm SM}_1\left(U(1)^3_{B-L} \right) + \mathcal{A}^{\rm New}_1\left(U(1)^3_{B-L} \right) \nonumber \\
\hspace*{2cm}  &&=-3 +\left [\left(\frac{1}{3}\right) ^3 
    +\left(\frac{2}{3}\right) ^3
    +\left(\frac{2}{3}\right) ^3 + \left(\frac{4}{3}\right) ^3 
    \right ]=0\, ,\nonumber \\
\mathcal{A}_2\left[\mbox{gravity}^2 \times U(1)_{B-L} \right] &&\propto   
     \mathcal{A}^{\rm SM}_2\left(U(1)_{B-L} \right) + \mathcal{A}^{\rm New}_2\left(U(1)_{B-L} \right) \nonumber \\
\hspace*{2cm}     &&=-3 + \left [\left(\frac{1}{3}\right) 
    +\left(\frac{2}{3}\right) 
    +\left(\frac{2}{3}\right) + \left(\frac{4}{3}\right)\right]=0\, . \nonumber 
\end{eqnarray}

We study scalar dark matter in an uncomplicated $\rm U(1)_{B-L}$ gauge extension of SM. Apart from the existing SM particle content, as mentioned earlier four exotic fermions ($N_{iR}$'s, where $i=1,2,3,4$), assigned with fractional $\rm B-L$ charges $-1/3,-2/3, -2/3$ and $-4/3$ are added to avoid the unwanted triangle gauge anomalies.  We add three scalar singlets in the process of breaking ${\rm B-L}$ gauge symmetry spontaneously, with two of them i.e., $\phi_1$, $\phi_2$ generate mass terms to the new fermions and $\phi_3$ helps in generating neutrino mass by type-I seesaw. An inert scalar singlet $\phi_{\rm DM}$, qualifies as a dark matter in the present model, whose stability is ensured by the $Z_2$ symmetry. The complete field content along with their corresponding charges under $\rm SU(2)_L\times U(1)_Y\times U(1)_{B-L}\times Z_2$ are provided in Table  \ref{model_charges}\,. 

%================================================
\begin{table}[tb!]
\begin{center}
%================================================
\begin{tabular}{|c|c|c|c|c|}
	\hline
		& Field	& $\rm SU(2)_L\times U(1)_Y$	& $\rm U(1)_{B-L}$ & $\rm Z_2$	\\
	\hline
	\hline
	Fermions& $Q_L \equiv(u, d)^T_L$		& $(\textbf{2}, 1/6)$	& $1/3$	 & $+$\\
		& $u_R$					& $(\textbf{1}, 2/3)$	& $1/3$ & $+$	\\
		& $d_R$					& $(\textbf{1}, -1/3)$	&$1/3$	& $+$\\
		& $\ell_L \equiv(\nu,~e)^T_L$		& $(\textbf{2}, -1/2)$	&  $-1$ & $+$	\\
		& $e_R$					    & $(\textbf{1}, -1)$ &  $-1$ & $+$	\\ \cline{2-5}
       
		& $N_{1R}$				    & $(\textbf{1}, 0)$	&   $-1/3$  & $+$	\\
		& $N_{2R}$				    & $(\textbf{1}, 0)$	&   $-2/3$	& $+$\\
		& $N_{3R}$				    & $(\textbf{1}, 0)$	&   $-2/3$	& $+$\\
		& $N_{4R}$				    & $(\textbf{1}, 0)$	&   $-4/3$	& $+$\\
     	\hline
	Scalars	& $H$					& $(\textbf{2}, 1/2)$	&   $0$	 & $+$\\
			& $\phi_{\rm DM}$			& $(\textbf{1}, 0)$	&   $-1/3$ & $-$	\\
			& $\phi_{1}$		& $(\textbf{1}, 0)$	&   $1$  & $+$	\\  
			& $\phi_{2}$			& $(\textbf{1}, 0)$	&   $2$	  & $+$	\\
			&$\phi_3$	      &$(\textbf{1},  0)$    &   $-1/3$   & $+$   \\
	%		&$\Delta$	      &$(\textbf{3},  2)$    &   $-2/3$   & $+$   \\
%			 & $\eta $ & 	 $(\textbf{2}, 1/2)$ & $-3 $\\
%			& $\Delta $  		& $(\textbf{3}, 1)$ & $-6 $\\
	\hline
	\hline
\end{tabular}
\caption{Particle spectrum and their charges of the proposed $\rm U(1)_{B-L}$ model.}
\label{model_charges}
\end{center}
\end{table}
%================================================
The relevant terms in the fermion interaction Lagrangian is given by 
\begin{eqnarray}  \label{Lagrangian}
&&\mathcal{L}^{\rm fermion}_{\rm Kin.}=
      \overline{Q}_L i \gamma^\mu \left(\partial_\mu+ig \frac{\vec{\tau}}{2} \cdot \vec{W}_\mu 
      +   \frac{1}{6} i\,g^\prime \,B_\mu + \frac{1}{3} i\,g_\text{BL} \,Z_\mu^\prime\right) Q_L   \nonumber \\
&&+  \overline{u_{R}} i \gamma^\mu \left(\partial_\mu
      +   \frac{2}{3} i \,g^\prime \,B_\mu + \frac{1}{3} i\,g_\text{BL} \,Z_\mu^\prime \right) u_{R} \nonumber \\
&&+  \overline{d_{R}} i \gamma^\mu \left(\partial_\mu
      -   \frac{1}{3} i\,g^\prime \,B_\mu + \frac{1}{3} i\,g_\text{BL} \,Z_\mu^\prime \right) d_{R} \nonumber \\
&&+  \overline{\ell_{L}} i \gamma^\mu \left(\partial_\mu+i g \frac{\vec{\tau}}{2} \cdot \vec{W}_\mu 
      -   \frac{1}{2} i\,g^\prime \,B_\mu - i\,g_\text{BL} \,Z_\mu^\prime \right) \ell_{L}            \nonumber \\
    &&+ \overline{e_{R}} i \gamma^\mu \left(\partial_\mu
      -  i\,g^\prime \,B_\mu -i\,g_\text{BL} \,Z_\mu^\prime  \right) e_{R}      \nonumber \\
&&+ \overline{N_{1R}} i \gamma^\mu \left(\partial_\mu
       -\left(\frac{1}{3}\right) i\,g_\text{BL} \,Z_\mu^\prime \right) N_{1R} 
       + \overline{N_{2R}} i \gamma^\mu \left(\partial_\mu
       -\left(\frac{2}{3}\right) i\,g_\text{BL} \,Z_\mu^\prime \right) N_{2R}\nonumber \\
&&+ \overline{N_{3R}} i \gamma^\mu \left(\partial_\mu
       -\left(\frac{2}{3}\right) i\,g_\text{BL} \,Z_\mu^\prime \right) N_{3R}\;+\overline{N_{4R}} i \gamma^\mu \left(\partial_\mu
       -\left(\frac{4}{3}\right) i\,g_\text{BL} \,Z_\mu^\prime \right) N_{4R}\;.
\end{eqnarray}
The Yukawa interaction for the present model is given by
\begin{eqnarray}
\mathcal{L}_{\rm Yuk}&=& -Y_u\, \overline{Q_{L}} \widetilde{H} u_{R} - Y_d \overline{Q_{L}}  H\, d_{R} -Y_e\, \overline{\ell_{L}} H e_{R}+ {\rm H.c} \nonumber \\
  &-& \left(\sum_\alpha \sum_{\beta = 2,3 } \frac{Y^\prime_{\alpha \beta}}{\Lambda} \overline{\ell_{\alpha L}} \tilde{H} N_{\beta R} \phi_3+ {\rm{H.c}}\right)-\sum_\alpha \left(\frac{Y^\prime_{\alpha 4}}{\Lambda} \overline{\ell_{\alpha L}} \tilde{H} N_{4 R} \phi^\dagger_3+ {\rm{H.c}}\right)\nn\\
  &-&  \sum_{\beta=2,3} \left(h_{\beta 1} \phi_1 \overline{N^c_{\beta R}} N_{1R}  + h_{\beta 4}\phi_2 \overline{N^c_{\beta R}}  N_{4 R}+~{\rm H.c} \right) -h^\prime_{11} \left(\overline{N^c_{1R}} N_{1R} \left(\frac{{\phi^\dagger_3}^2+\phi_1 \phi_3}{\Lambda}\right) \right),\nn \\  
  &-& \frac{h^\prime_{22}}{\Lambda} \overline{N_{2R}^c} N_{2R} \phi_1 \phi_3^\dagger- \frac{h^\prime_{33}}{\Lambda} \overline{N_{3R}^c} N_{3R} \phi_1 \phi_3^\dagger - \frac{ h^\prime_{23}}{\Lambda}\overline{N_{2R}^c} N_{3R} \phi_1 \phi_3^\dagger -\frac{ h^\prime_{14}}{\Lambda}\overline{N_{1R}^c} N_{4R} \phi_2 \phi_3 \nn \\
  &-& \frac{h^\prime_{34}}{\Lambda}\overline{N_{3R}^c} N_{4R} \phi_1^2-\frac{h^\prime_{24}}{\Lambda} \overline{N_{2R}^c} N_{4R} \phi_1^2+{\rm ~H.c}.
\label{dirac}
\end{eqnarray}
with $\widetilde{H}=i\sigma_2 H^*$. The interaction Lagrangian for the scalar sector is as follows
\begin{eqnarray}
\mathcal{L}^{\rm }_{\rm scalar} &=&
      \left(\mathcal{D}_\mu H \right)^\dagger \left(\mathcal{D}^\mu H\right) + \left(\mathcal{D}_\mu \phi_{\rm DM} \right)^\dagger \left(\mathcal{D}^\mu \phi_{\rm DM}\right)
      +\left(\mathcal{D}_\mu \phi_{1}\right)^\dagger \left(\mathcal{D}^\mu \phi_{1}\right)\nonumber \\
      &&+\left(\mathcal{D}_\mu \phi_{2}\right)^\dagger \left(\mathcal{D}^\mu \phi_{2}\right)+ \left(\mathcal{D}_\mu \phi_{3}\right)^\dagger \left(\mathcal{D}^\mu \phi_{3}\right)
      +V\left(H,\phi_{\rm DM},\phi_1,\phi_2, \phi_3 \right),
\end{eqnarray} 
%+{\rm Tr}\left[\left(\mathcal{D}_\mu \Delta \right)^\dagger \left(\mathcal{D}^\mu \Delta\right)\right]
where the covariant derivatives are
\begin{eqnarray} 
&&\mathcal{D}_\mu H = \partial_{\mu} H+i\,g \vec{W}_{\mu L}\cdot \frac{\vec{\tau}}{2}\, H  \,+\, i\frac{g^{\prime}}{2}B_{\mu} H\, , \nonumber \\
&&\mathcal{D}_\mu \phi_{\rm DM} =\partial_{\mu} \phi_{\rm DM} -\frac{1}{3}i g_{\rm BL} \,Z_\mu^\prime \phi_{\rm DM}  \, , \nonumber \\
&&\mathcal{D}_\mu \phi_{1} =\partial_{\mu} \phi_{1} +i g_{\rm BL} \,Z_\mu^\prime \phi_{1}  \, , \nonumber \\
&&\mathcal{D}_\mu \phi_{2} =\partial_{\mu} \phi_{2} + 2 i g_{\rm BL} \,Z_\mu^\prime \phi_{2}\, , \nonumber\\
&&\mathcal{D}_\mu \phi_{3} =\partial_{\mu} \phi_{3} - \frac{1}{3} i g_{\rm BL} \,Z_\mu^\prime \phi_{3}.
%&&\mathcal{D}_\mu \Delta = \partial_{\mu} \Delta + i\,g \vec{W}_{\mu a}\cdot \frac{\vec{\tau_a}}{2}\, \Delta \,+\, ig^{\prime} B_{\mu a} \Delta-\frac{2}{3} i g_{\rm BL} \,Z_\mu^\prime \Delta\,,
\end{eqnarray}
And the scalar potential takes the form
\begin{eqnarray}
V^\prime&=& \mu^2_{\rm H}  H^\dagger H + \lambda_{\rm H} (H^\dagger H)^2 + \mu^2_1 \phi^\dagger_1 \phi_1 + \lambda_1 (\phi^\dagger_1 \phi_1)^2 + \mu^2_2 \phi^\dagger_2 \phi_2  + \lambda_2 (\phi^\dagger_2 \phi_2)^2 \nonumber \\
     &&+ \mu^2_{\rm D} \phi_{\rm DM}^\dagger {\phi_{\rm DM}} + \lambda_{\rm D} (\phi_{\rm DM}^\dagger {\phi_{\rm DM}})^2  +\lambda_{\rm H1} (H^\dagger H) (\phi^\dagger_1 \phi_1)
      +\lambda_{\rm H2} (H^\dagger H) (\phi^\dagger_2 \phi_2)\nonumber \\
      &&+ \lambda_{12} (\phi^\dagger_1 \phi_1) (\phi^\dagger_2 \phi_2) + \lambda_{\rm DH} (H^\dagger H) (\phi_{\rm DM}^\dagger \phi_{\rm DM}) + \lambda_{\rm D1} (\phi_{\rm DM}^\dagger \phi_{\rm DM}) (\phi^\dagger_1 \phi_1) \nonumber \\
      &&+ \lambda_{\rm D2} (\phi_{\rm DM}^\dagger \phi_{\rm DM}) (\phi^\dagger_2 \phi_2) +\mu_{12} \left((\phi_1)^2 \phi_2^\dagger + (\phi_1^\dagger)^2 \phi_2\right),
     \label{scalar}
 \end{eqnarray} 
 \begin{eqnarray}
V^{\prime\prime} &=& \mu^2_3 \phi^\dagger_3 \phi_3  + \lambda_3 (\phi^\dagger_3 \phi_3)^2+\lambda_{\rm H3} (H^\dagger H) (\phi^\dagger_3 \phi_3)\nn \\
     &&+\lambda_{13} (\phi^\dagger_1 \phi_1) (\phi^\dagger_3 \phi_3)+\lambda_{23} (\phi^\dagger_2 \phi_2) (\phi^\dagger_3 \phi_3)+\lambda_{m}((\phi_3)^3\phi_1+ (\phi^\dagger_3)^3 \phi^\dagger_1)  \nn\\
 &&+\lambda_{D3} (\phi^\dagger_{\rm DM} \phi_{\rm DM}) (\phi^\dagger_3 \phi_3)+\lambda_{\rm DD}((\phi^\dagger_{\rm DM} \phi_3)^2+ (\phi^\dagger_3\phi_{\rm DM} )^2),\nn \\
     \label{scalar2}
 \end{eqnarray} 
Full potential of this model is given by
\begin{equation}
V(H,\phi_1,\phi_2,\phi_3,\phi_{\rm DM})=V^\prime+V^{\prime\prime}.\label{totpot}
\end{equation}
Here, $\phi_{\rm DM} = \frac{\rm S_{DM} + i A_{DM}}{\sqrt{2}}$ is the DM singlet in the present model. The stability of the potential is assured by the copositive criteria, given as  
\begin{eqnarray}
&& \lambda_{\rm H},\lambda_1,\lambda_2,\lambda_{\rm DM},\lambda_{\rm 3} \geq 0,\;\; \lambda_{\rm H1}+\sqrt{\lambda_{\rm H} \lambda_1} \geq 0, \nn \\
&& \lambda_{\rm H2}+\sqrt{\lambda_{\rm H} \lambda_2} \geq 0,\;\; \lambda_{\rm HD}+\sqrt{\lambda_{\rm H} \lambda_{\rm D}} \geq 0, \nn \\
&& \lambda_{\rm H3}+\sqrt{\lambda_{\rm H} \lambda_{\rm 3}} \geq 0,\;\; \lambda_{12}+\sqrt{\lambda_1 \lambda_2} \geq 0, \nn \\
&& \lambda_{\rm D1}+\sqrt{\lambda_{\rm D} \lambda_1} \geq 0, \;\; \lambda_{\rm D2}+\sqrt{\lambda_{\rm D} \lambda_2} \geq 0, \nn \\
&& \lambda_{\rm 13}+\sqrt{\lambda_{\rm 3} \lambda_1} \geq 0,\;\; \lambda_{\rm 23}+\sqrt{\lambda_{\rm 3} \lambda_2} \geq 0,\;\;\lambda_{\rm D3}+\sqrt{\lambda_{\rm D} \lambda_{ 3}} \geq 0.
\end{eqnarray}
%====================================================
\section{Spontaneous symmetry breaking and mixing}
%======================================================
Spontaneous symmetry breaking of $\rm SU(2)_L\times U(1)_Y \times U(1)_{B-L}$ to $\rm SU(2)_L\times U(1)_Y$ is realized by assigning non-zero vacuum expectation value (VEV) to the scalar singlets $\phi_1$, $\phi_2$ and $\phi_3$. Later, the SM gauge group gets spontaneously broken to low energy theory by the SM Higgs doublet $H$. The scalar sector can be written in terms of CP even and CP odd components as
\begin{align}
&H^0 =\frac{1}{\sqrt{2} }(v+h)+  \frac{i}{\sqrt{2} } A^0\,, \nonumber \\
& \phi_1 = \frac{1}{\sqrt{2} }(v_1+h_1)+  \frac{i}{\sqrt{2} } A_1\,, \nonumber \\
& \phi_2 = \frac{1}{\sqrt{2} }(v_2+h_2)+  \frac{i}{\sqrt{2} } A_2\,,\nn\\
& \phi_3 = \frac{1}{\sqrt{2} }(v_3+h_3)+  \frac{i}{\sqrt{2} } A_3\,,\nn
\end{align} 
where, $\langle H\rangle=(0, v/\sqrt2)^T$, $\langle \phi_1\rangle=v_1/\sqrt2$, $\langle \phi_2\rangle=v_2/\sqrt2$ and $\langle \phi_3\rangle=v_3/\sqrt2$.
% and the VEV of scalar triplet is given as
%$\langle \Delta \rangle = \frac{v_{\rm T}}{\sqrt{2}} \begin{pmatrix}0 & 0\\
%1 &0 \end{pmatrix}$. 
%========================================
\subsection{Mixing in scalar sector}
%============================================
The minimisation conditions of the scalar potential in Eq.\eqref{totpot} correspond to 
\begin{eqnarray}
 &&\mu^2_{\rm H} = -\frac{1}{2}\left[2 \lambda_H v^2 + \lambda_{\rm H1}{v^2_1} +  \lambda_{\rm H2}{v^2_2}+\lambda_{\rm H3}{v^2_{3}}\right] ,\nn\\
 &&\mu^2_{1} = -\frac{1}{2}\left[2 \lambda_{1} v^2_1 +\lambda_{\rm H1}{v^2} +\lambda_{12}{v^2_2} + \lambda_{\rm 13} v^2_3+ 2\sqrt{2} \mu_{12} v_2 + \frac{\lambda_m v^3_3}{v_1}\right] ,\nn\\
 &&\mu^2_{2} =  -\frac{1}{2}\left[2 \lambda_{2} v^2_2 + \lambda_{\rm H2} v^2 +\lambda_{12}{v^2_1} +\lambda_{\rm 23}{v^2_{3}}+ \sqrt{2} \mu_{12} \frac{{v_1}^2}{v_2}\right] ,\nn\\
% &&\mu^2_{\rm T} = -\frac{1}{2}\left[2 \lambda_{\rm T1} v^2_1 + \lambda_{\rm T2}{v^2_2} +  \lambda_{\rm HT}{v}^2+2\lambda_{\rm T}{v^2_{\rm T}}\right],\nn \\
 && \mu^2_3 =-\frac{1}{2}\left[2 \lambda_3 v^2_3 + \lambda_{\rm H3} v^2 +\lambda_{13}v^2_1 + \lambda_{23}v^2_2 +3 \lambda_m v_1 v_3 \right].
\end{eqnarray}
%The neutral scalar mass matrix constructed from the singlet Higgs fields is given by
%\begin{equation}
%M^2_{12}=\begin{pmatrix}
% 2\lambda_{1} v^2_1 & v_1(\lambda_{12}v_2+\sqrt{2} \mu_{12})  \\
% v_1(\lambda_{12}v_2+\sqrt{2} \mu_{12}) & 2\lambda_{2} v^2_2-\frac{\mu_{12} v^2_1}{\sqrt{2} v_2}
% \end{pmatrix} 
%  \end{equation}
%After the electroweak symmetry breaking, the CP-even scalar mass matrix takes the form
%\begin{equation}
%M^2_{E}=\begin{pmatrix}
%2\lambda_{\rm H} v^2 & \lambda_{\rm H1} v_1 v & \lambda_{\rm H2} v_2 v & \lambda_{H3} v_3 v \\
%\lambda_{\rm H1} v_1 v & 2\lambda_{1} v^2_1 -\frac{\lambda_m v^3_3}{2 v_1} & v_1(\lambda_{12}v_2+\sqrt{2} \mu_{12}) & \lambda_{13} v_1 v_3 +\frac{3 \lambda_m v^2_3}{2}  \\
% \lambda_{\rm H2} v_2 v & v_1(\lambda_{12}v_2+\sqrt{2} \mu_{12}) & 2\lambda_{2} v^2_2-\frac{\mu_{12} v^2_1}{\sqrt{2} v_2} & \lambda_{23} v_2 v_3\\
% \lambda_{H3} v v_3 & \lambda_{13} v_1 v_3 +\frac{3 \lambda_m v^2_3}{2} & \lambda_{23} v_2 v_3 & \frac{1}{2} (3\lambda_m v_1 v_3 +4 \lambda_3 v^2_3)
% \end{pmatrix}. \label{higgsmix}
%  \end{equation}
We assume that the third neutral Higgs ($\phi_3$) is heavy, which leads to a small mixing and decoupled. Thus the mixing of CP even scalar fields in the flavor basis $(H,~\phi_1,~\phi_2)$ is 
\begin{eqnarray}
M^2_E =\begin{pmatrix}
2\lambda_{\rm H} v^2 & \lambda_{\rm H1} v_1 v & \lambda_{\rm H2} v_2 v \\
\lambda_{\rm H1} v_1 v & 2\lambda_{1} v^2_1 -\frac{\lambda_m v^3_3}{2 v_1} & v_1(\lambda_{12}v_2+\sqrt{2} \mu_{12})\\
 \lambda_{\rm H2} v_2 v & v_1(\lambda_{12}v_2+\sqrt{2} \mu_{12}) & 2\lambda_{2} v^2_2-\frac{\mu_{12} v^2_1}{\sqrt{2} v_2}
 \end{pmatrix}.
\end{eqnarray}
For simplicity in diagonalizing the above mass matrix, we assume $\lambda_{H1} < \lambda_H$, $\lambda_{H1}=\lambda_{H2}$ and $v_1=v_2$. Thus we obtain an equivalent form of the above mass matrix as follows
\begin{equation}
M^2_E \simeq \begin{pmatrix}
a && a && a\\
a && y && b\\
a && b && y
\end{pmatrix}.
\end{equation}
The matrix that diagonalizes the CP even mass matrix in the limit of minimal mixing with SM Higgs, is given by
\begin{equation}
U_E =\begin{pmatrix}
1 && \beta \cos{\alpha}-\beta \sin{\alpha} && \beta \cos{\alpha}+\beta \sin{\alpha}\\
-\beta && \cos{\alpha} && \sin{\alpha}\\
-\beta && -\sin{\alpha} && \cos{\alpha}
\end{pmatrix},
\end{equation}
here $\alpha$ denotes the mixing between $\phi_1$ and $\phi_2$, and $\beta$ represents the mixing of Higgs with rest two scalars. For $\alpha = \frac{5\pi}{4}$ and $\beta =\frac{a}{b+y-a}$, one can go the mass eigen basis $(H^\prime,~H_1,~H_2)$ with the mass eigenvalues 
\begin{eqnarray}
&& M^2_{H^\prime}= a- 4a\beta + 2(b+y)\beta^2,\nn\\
&& M^2_{H_1}= -b+y,\nn\\
&& M^2_{H_2}=b+y+2a\beta(2+\beta).
\end{eqnarray}
Solving for the parameters $a, b, y$, one can have
\begin{equation}
\beta=\frac{-M^2_{H^\prime}+M^2_{H_2}-\sqrt{-15 M^4_{H^\prime}-10 M^2_{H_2}M^2_{H^\prime}+M^4_{H_2}}}{4(2M^2_{H^\prime}+M^2_{H_2})}.
\end{equation}
The mass eigenstate $H^\prime$ is considered to be observed Higgs at LHC ($M_{\rm H^\prime} = 125$ GeV). This mixing is taken to be minimal ($\beta < 0.1$), such that it does not violate the LHC bounds on the observed Higgs. 
The mixing matrix of the CP odd sector in the basis ($A_1,~A_2,~A_3$) is given by 
 \begin{eqnarray}
 M^2_{O}=\begin{pmatrix}
 -2\sqrt{2}\mu_{12} v_2-\frac{\lambda_m v^3_3}{2 v_1} && \sqrt{2}\mu_{12} v_1 && -\frac{3}{2} \lambda_m v^2_3\\
 \sqrt{2}\mu_{12} v_1 && -\frac{\mu_{12} v^2_1}{\sqrt{2} v_2} && 0\\
 -\frac{3}{2} \lambda_m v^2_3 && 0 && -\frac{9}{2} \lambda_m v_1 v_3
  \end{pmatrix}.
 \end{eqnarray}
 To simplify the diagonalization, we assume $\lambda_{m}=\frac{\mu_{12} v_1}{v^2_2}$ and the above mass matrix gives two massive CP odd eigenstates ($A^\prime_1, A^\prime_2$) with masses $M^2_{A^\prime_1} \approx  3\mu_{12}v_2$ and $ M^2_{A^\prime_2}\approx 6 \mu_{12} v_2$ respectively. The third eigenstate, $A^\prime_3$ remains massless and gets absorbed by the new gauge boson $Z'$, acquiring the mass $M_{Z^\prime}= g_{\rm BL}\sqrt{v^2_1+4 v^2_2+\frac{1}{9} v^2_3}$.
%andmassive CP odd scalar will have a mass term $M_{Acp}=\frac{\mu_{12}({v_1}^2+4{v_2}^2)}{\sqrt{2}v_2}$.
%The longitudenal modes from the linear combination of CP odd scalars is given by
%\begin{eqnarray}
%&& A_{G}=-\frac{2 v_2}{{v_1}^2+4{v_2}^2}A_2+\frac{ v_1}{{v_1}^2+4{v_2}^2}A_1\\
%&& A_{CP}=\frac{ v_1}{{v_1}^2+4{v_2}^2}A_2+\frac{2 v_2}{{v_1}^2+4{v_2}^2}A_1\\
%\end{eqnarray}
\subsection{Comments on neutrino mass }
 We can have a tree level Dirac mass for the active neutrinos, which can be constructed from the 5-dimension Yukawa coupling in Eq.\eqref{dirac}. Therefore the tree level small neutrino Majorana mass matrix within type I seesaw framework can be obtained as
\begin{equation}
m_\nu = M_D M^{-1}_R M^T_D,
\end{equation}
where,
 \begin{equation}
M_D =\frac{v v_3}{2 \Lambda} \begin{pmatrix}
0 & Y^\prime_{12} & Y^\prime_{13} & Y^\prime_{14} \\
0 & Y^\prime_{22} & Y^\prime_{23} & Y^\prime_{24}\\
0 & Y^\prime_{32} & Y^\prime_{33} & Y^\prime_{34}
\end{pmatrix}.
\label{fermionmatrix2}
\end{equation}
\begin{figure}[t!]
\begin{center}
\includegraphics[height=55mm,width=70mm]{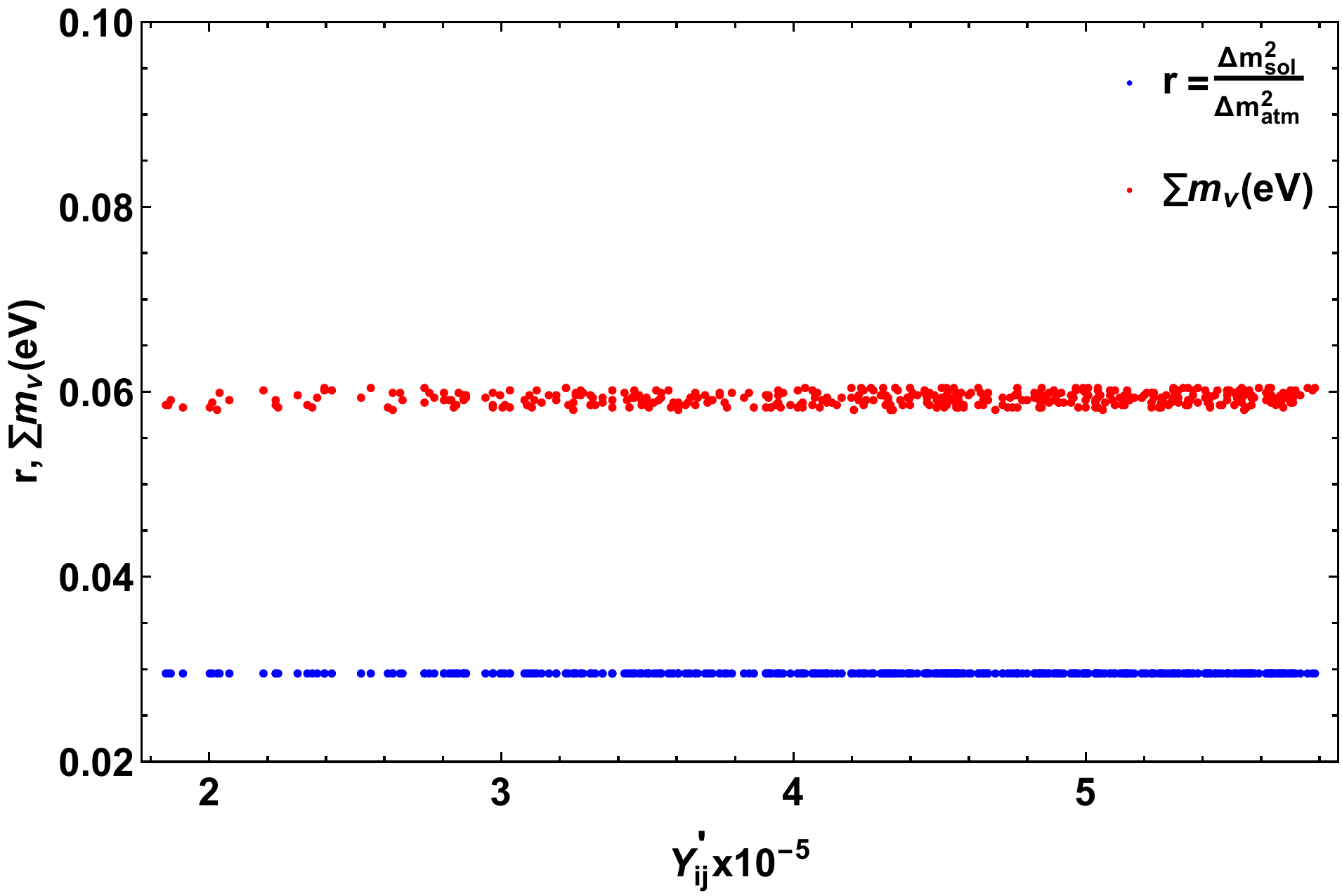}
\caption{Variation of Yukawa coupling with the sum of observed active neutrino masses(red) and ratio of solar to atmospheric mass squared difference (blue) allowed by the $3\sigma$ neutrino oscillation data \cite{Esteban:2020cvm,Aghanim:2018eyx}.}
\label{ynu_mnu_plot}
\end{center}
\end{figure}
The flavor structure of right-handed neutrino mass $M_R$ can be written from the Lagrangian in Eq.\eqref{dirac} as
\begin{equation}
M_R = \frac{1}{\sqrt{2}}\begin{pmatrix}
0 & h_{21}v_1 & h_{31}v_1 & 0\\
h_{21}v_1 & 0 & 0 & h_{24}v_2 \\
h_{31}v_1 & 0 & 0 & h_{34} v_2 \\
0 & h_{24}v_2 & h_{34}v_2  & 0 
\end{pmatrix}+\frac{1}{2 \Lambda} \begin{pmatrix}
h^\prime_{11} (v^2_3+ v_1 v_3) &0  & 0 & h^\prime_{14} v_2 v_3\\
0 & h^\prime_{22} v_1 v_3 & h^\prime_{23} v_1 v_3 & h^\prime_{24} v^2_1 \\
0 & h^\prime_{23} v_1 v_3 & h^\prime_{33} v_1 v_3 & h^\prime_{34} v^2_1 \\
h^\prime_{14} v_2 v_3 & h^\prime_{24} v^2_1 & h^\prime_{34} v^2_1  & 0 
\end{pmatrix}\,.
\label{fermionmatrix1}
\end{equation} 
The second term in the right hand side of the above equation corresponds to higher dimension corrections. To diagonalize this mass matrix, we take the simplistic assumption that all the couplings are of the same order and the correction terms are suppressed by the factor $\frac{v_{k}}{\Lambda} \approx 0.01$, where $k=1,2,3$. We found that the masses ($M_{i}$) of the heavy fermion eigenstates ($N_{D_i}$) as $0.012m$, $0.015m$, $3.13m$, $3.2m$, where $i=1,2,3,4$ and $m$ is the free mass parameter, which can adjusted to achieve the required order of Majorana mass for the heavy neutrinos. The corresponding eigenvector matrix can be obtained as follows
\begin{eqnarray}
U_{N}=\begin{pmatrix}
0.0083 && -0.7070 && 0.4998 && 0.5\\
-0.8944 && -0.0059 && -0.3129 && 0.3194\\
 0.4471 && 0.0047 && -0.6330 && 0.6319\\
 -0.002 &&  0.7070 && 0.5014 && 0.4985
\end{pmatrix}.
\end{eqnarray} 
We consider the Yukawa couplings to be complex and all are of similar order in magnitude.  With the right-handed neutrino masses in the range of $1$ to $100$ TeV, we represent the allowed regime of Yukawa in Fig.\ref{ynu_mnu_plot}, which satisfy the $3\sigma$ limit of neutrino oscillation data. 
\section{Phenomenology of singlet scalar Dark Matter}
\label{sec2}

\subsection{Relic density}

\begin{figure}[t!]
\begin{center}
\includegraphics[width=0.3\linewidth]{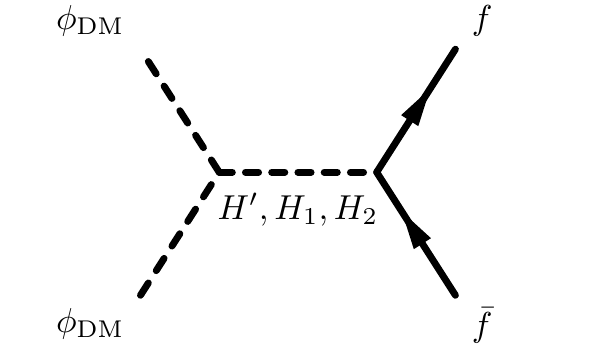}
\includegraphics[width=0.3\linewidth]{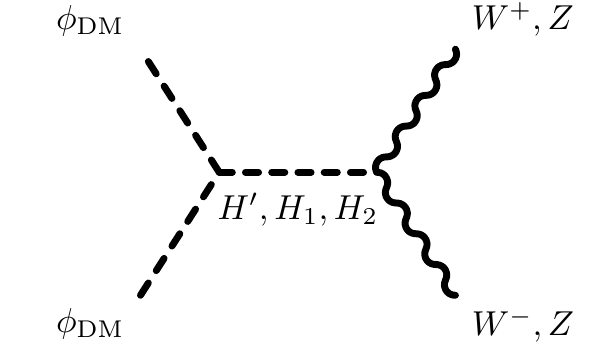}
\includegraphics[width=0.3\linewidth]{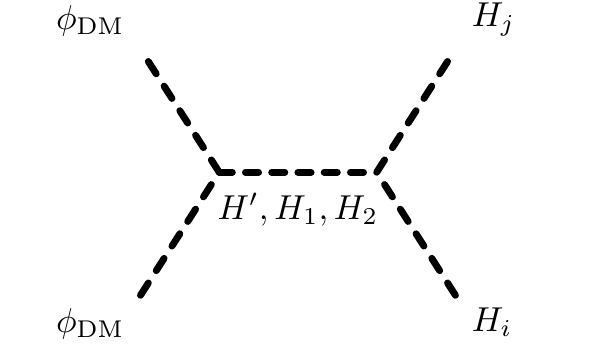}
\includegraphics[width=0.3\linewidth]{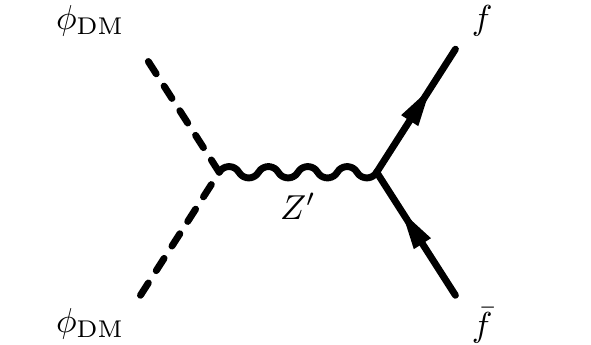}
\caption{Feynman diagrams those contribute to the relic density of DM in scalar and gauge portal portal.}
\label{feyndiag}
\end{center}
\end{figure}
       
\begin{figure}
\includegraphics[height=55mm,width=75mm]{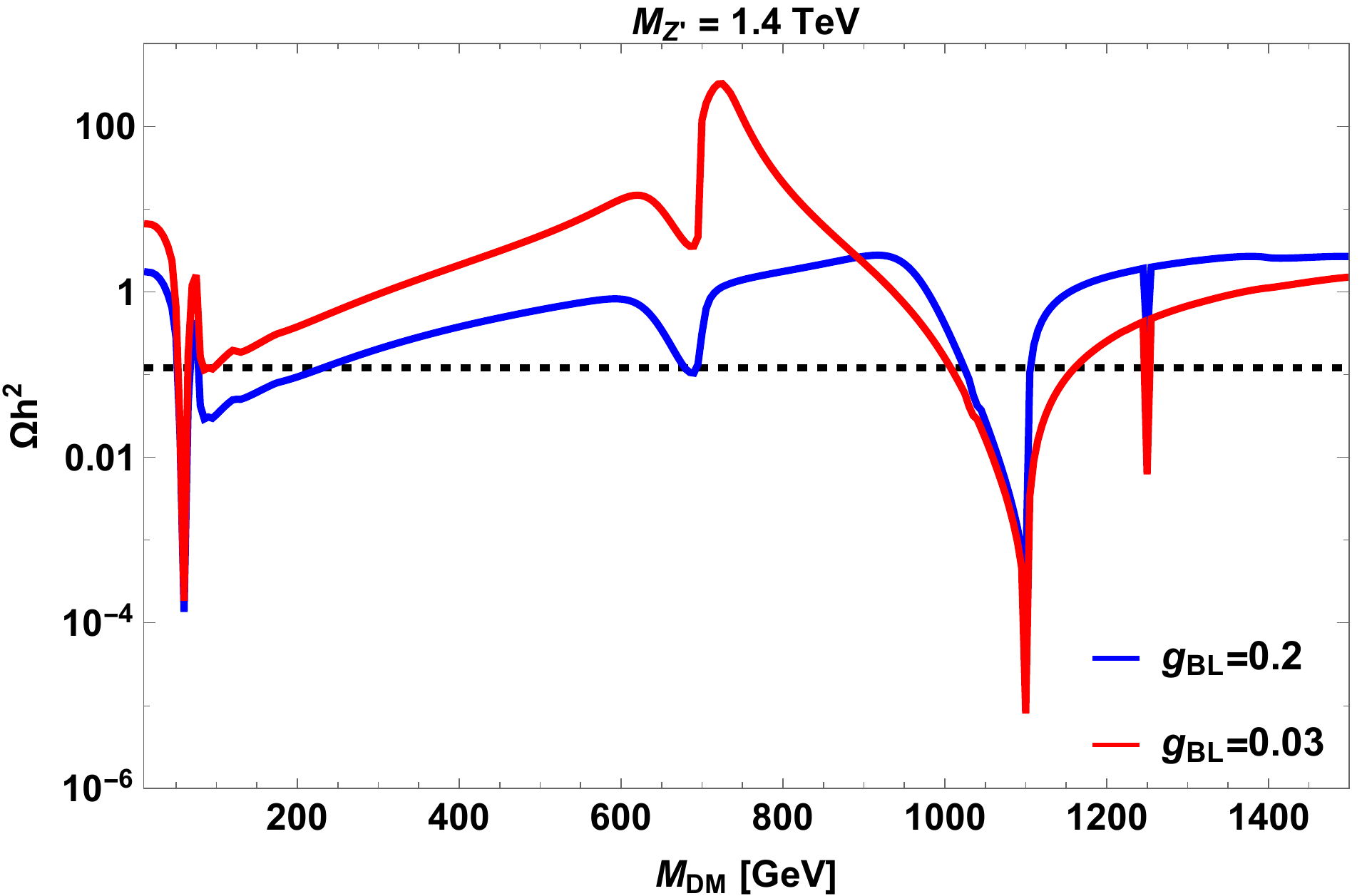}
\includegraphics[height=55mm,width=75mm]{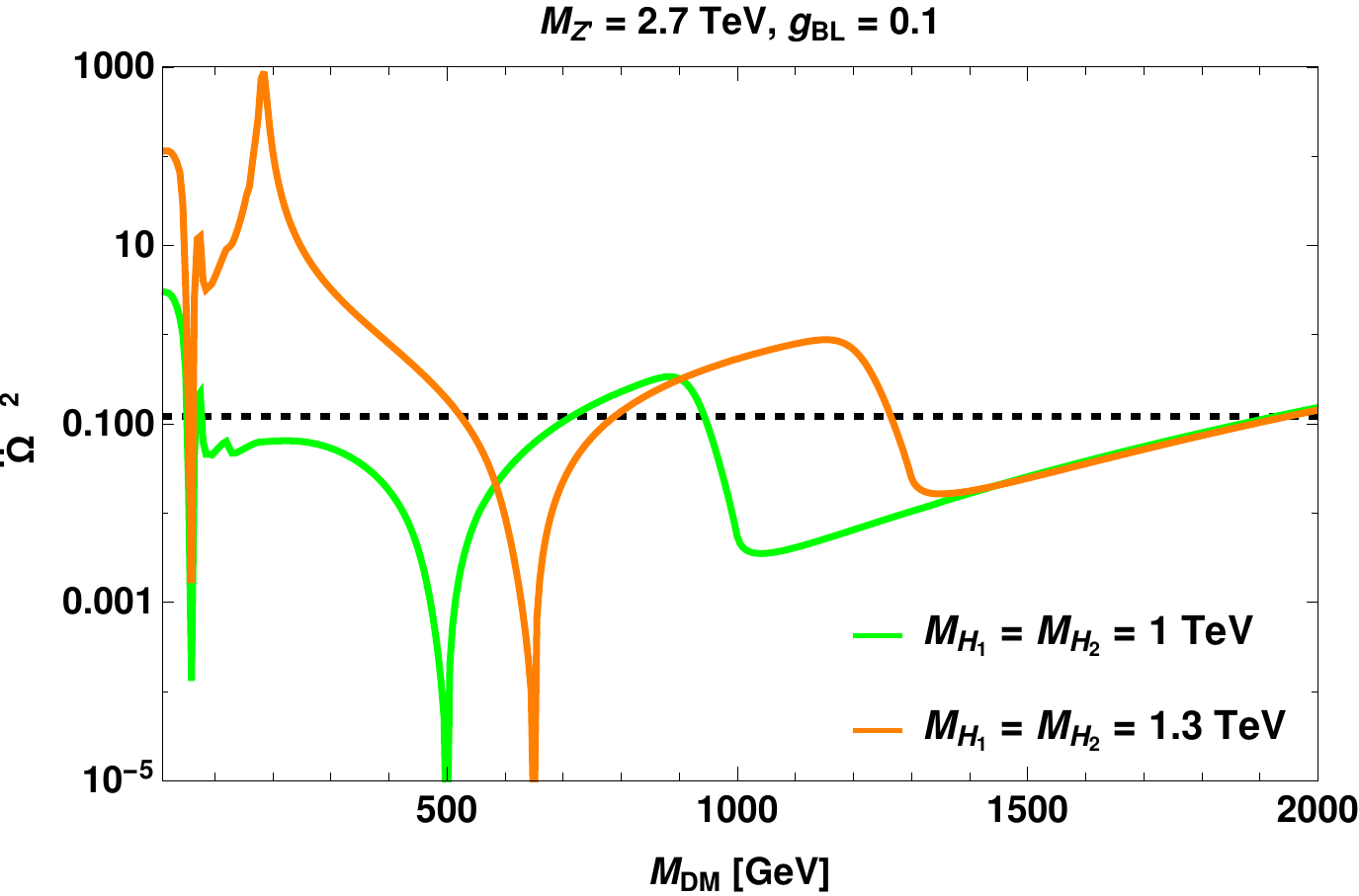}
\caption{Variation of relic density with DM mass for different values of gauge coupling $g_{\rm BL}$ (left panel) and $H_{1,2}$ mass (right panel). Horizontal dashed lines stand for Planck $3\sigma$ band. The scalar-DM coupling is taken to be $0.1$}
\label{relicv2}
\end{figure}

The model accommodates scalar DM, which has both scalar and gauge portal annihilation channels, provided in Fig. \ref{feyndiag}\,. The mass difference between the CP-odd and CP-even component of the singlet scalar is generated by $\phi_3$. Thus in addition to annihilations, co-annihilation channels also contribute to the DM relic density \cite{Griest:1990kh, Edsjo:1997bg, Bell:2013wua}. We use LanHEP\cite{Semenov:1996es} and micrOMEGAs \cite{Pukhov:1999gg, Belanger:2006is, Belanger:2008sj} for the model implementation and DM analysis. %To illustrate the behavior of relic density, we fix the mass spectrum \blue{($M_{Z'}, M_{H_1}, M_{H_2}$) as ($1.4,2.2,2.5$) TeV.} 
Fig. \ref{relicv2} depicts relic density as a function of DM mass for various set of values for model parameters.  We found that for DM mass below 75 GeV, the annihilation to fermion anti-fermion pair (except $t\bar{t}$) maximally contribute to relic density, both in scalar and gauge portals. Once kinematically allowed, relic density gets contribution from the channels with SM gauge bosons, $t\bar{t}$ and scalar bosons in the final state as well. Because of s-channel annihilations, the resonances (dips) are observed when the DM mass gets closer to half of the propogator mass. With ($M_{Z'}, M_{H_1}, M_{H_2}$) = ($1.4,2.2,2.5$) TeV, left panel of Fig. \ref{relicv2} shows the behavior of DM abundance for two set of values for $g_{\rm BL}$. For lower gauge coupling (red), only scalar mediated channels  contribute to relic density. While for large gauge coupling (blue), cross sections of $Z^\prime$-portal channels also add up, hence the curve takes a  plateau shape. Right panel projects the shift in scalar resonance ($H_1,H_2$) according to their mass, where we took $g_{\rm BL} = 0.1$ and $M_{Z^\prime} = 2.7$ TeV. In this case, the $Z^\prime$ mediated contribution is minimal with large propagator suppression, while scalar portal channels contribute maximally. The decrease in relic density after $1$ TeV (green) and $1.3$ TeV (orange) is due to contributions from the channels with $H_1 H_1$, $H_2 H_2$ in the final state via scalar propagator. 

\subsection{Direct searches}
\begin{figure}[t!]
\begin{center}
\includegraphics[width=0.3\linewidth]{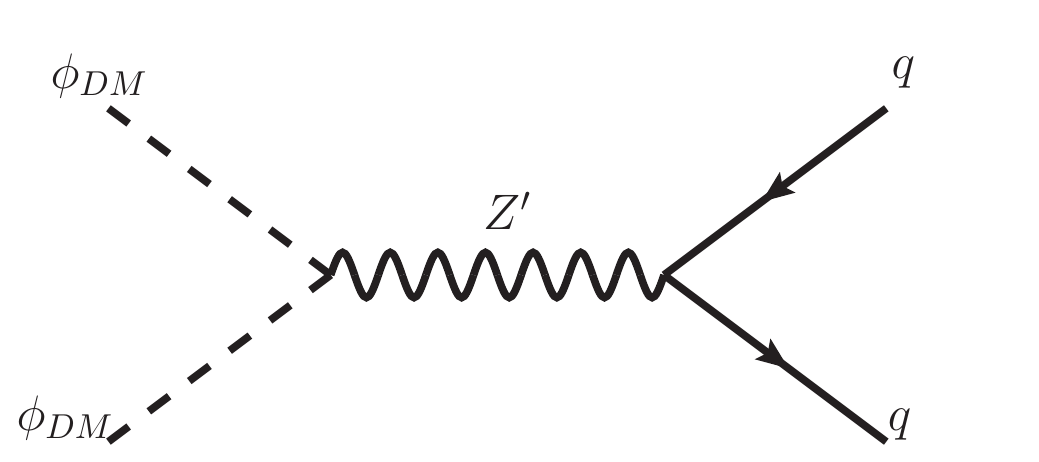}
\caption{t-channel scattering of DM with nucleus.}\label{ddfeyn}
\end{center}
\end{figure}

Now we look for the constraints on the model parameters due to direct detection limits. The effective Lagrangian for $Z^{\prime}$-mediated t-channel process shown in Fig. \ref{ddfeyn}, is  given as
\bea%
\mathcal{L}^{V}_{\mathrm{eff}}\supset
-\frac{{n_{\rm DM}} g_{\rm BL}^2}{3 M_{Z^{\prime}}^2}
\left(S\partial^{\mu}A-A\partial^{\mu}S  \right)\bar{u}\gamma_\mu u
-\frac{{n_{\rm DM}}g_{\rm BL}^2}{3 M_{Z^{\prime}}^2}
\left(S\partial^{\mu}A-A\partial^{\mu}S  \right) \bar{d}\gamma_\mu d\,.%
\eea\label{ssdmBL_eff1} %
%Comparing with Eqn. (\ref{dm_ddvector_lag}), one can find the value of $b_{p,n}$ as
%\be
%b_p=b_n=\frac{{n_{\rm DM}} g_{\rm BL}^2}{M_{Z^{\prime}}^2}. \nn%
%\ee%
%From Eqn. (\ref{dm_ddvector_cross}), the SI WIMP-nucleon contribution is given by
The corresponding spin independent (SI) WIMP-nucleon cross section turns out to be
\be
\sigma^{Z^{\prime}}_{\rm SI}=\frac{{\mu}^2}{\pi}\frac{{n^2_{\rm DM}} g_{\rm BL}^4}{M_{Z^{\prime}}^4}\,.
\ee
Here, $\mu$ denotes the reduced mass of DM-nucleon system. The t-channel scalar exchange i.e via $H^\prime$, $H_1, H_2$, can also give a SI contribution, but this is not relevant for the purpose of our study.
\subsection{Collider constraints}
\begin{figure}[t!]
\begin{center}
\includegraphics[width=0.48\linewidth]{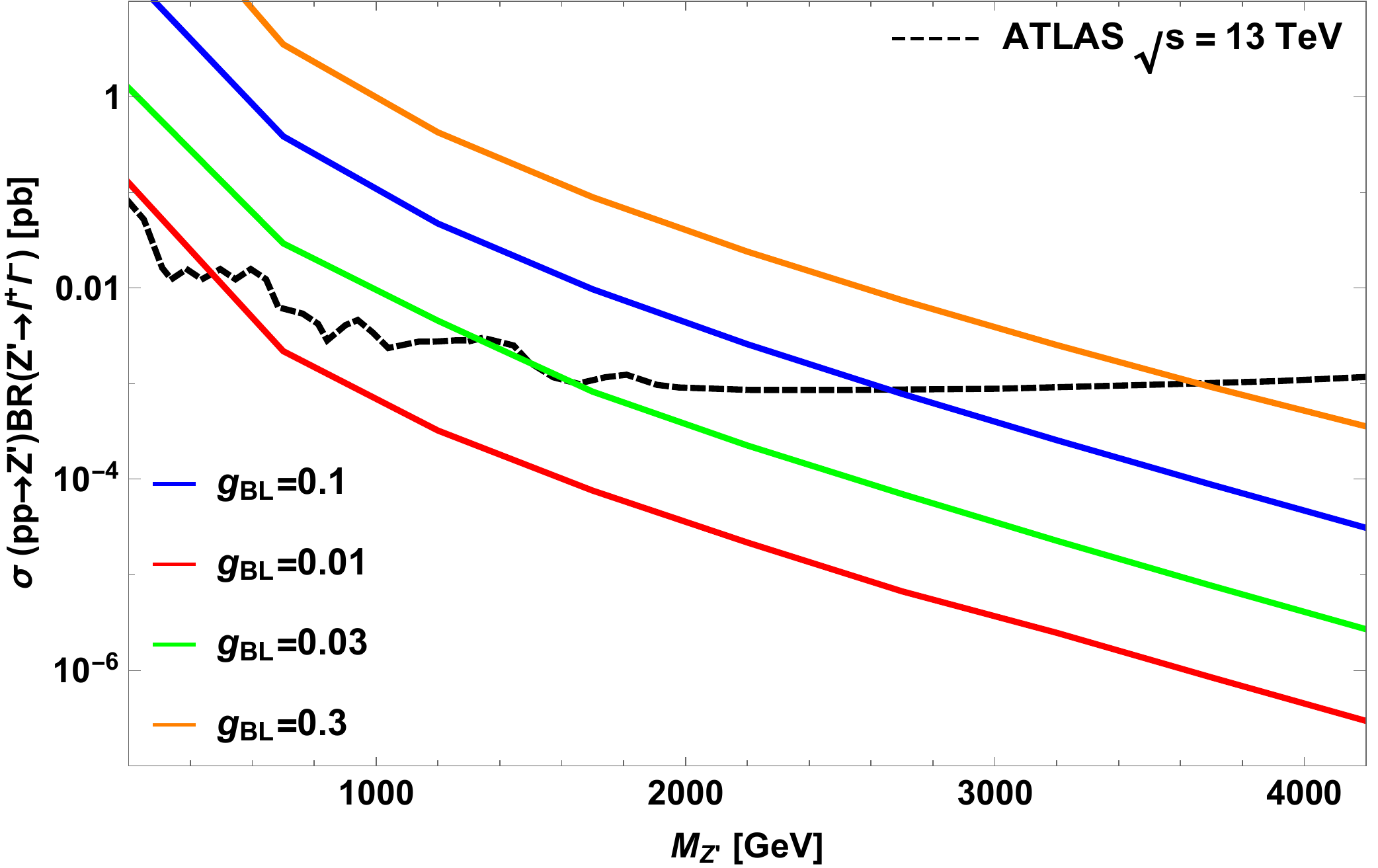}
\caption{Colored lines represent the dilepton signal cross section as a funciton of $M_{Z^\prime}$ for different values of $ g_{\rm BL}$ with the black dashed line points to ATLAS bound \cite{TheATLAScollaboration:2015jgi}.} \label{collidermz}
\end{center}
\end{figure}

ATLAS and CMS experiments are searching for new heavy resonances in both dilepton and dijet signals. It is found in the recent past, that these two experiments provide lower limit on $Z^\prime$ boson with dilepton signature, resulting a stronger bound than dijets due to relatively fewer background events. The investigation for $Z^\prime$, through dilepton signals from ATLAS experiment \cite{TheATLAScollaboration:2015jgi} concluded with stringent limit on the ratio of $Z^\prime$ mass ($M_{Z^\prime}$) and the gauge coupling ($ g_{\rm BL}$).
  We use CalcHEP \cite{Belyaev:2012qa,Kong:2012vg} to calculate the production cross section of $Z^\prime$ to dilepton ($e^+ e^-$, $\mu^+,\mu^-$) in final states. The variation of $Z'$ production cross section times the branching of dilepton as a function of $M_{Z'}$ is shown in Fig. \ref{collidermz}. From this plot we can interpret that, for $ g_{\rm BL}=0.01$, $M_{Z^\prime} < 0.5$ TeV regime is excluded by the ATLAS bound. Similarly for $g_{\rm BL}=0.03$, the mass regime for $M_{Z'} < 1.4$ TeV is not allowed. We found with a little larger values of $ g_{\rm BL}=0.1,0.3$, the allowed mass regime for $M_{Z'}$ should be greater than $2.7$ TeV and $3.7$ TeV,  respectively. Furthermore, there is also a lower limit on the ratio $\frac{M_{Z^\prime}}{g_{\rm BL}}$ from LEP-II \cite{Schael:2013ita}, i.e., $6.9$ TeV.

\begin{figure}[h!]
\begin{center}
\includegraphics[width=0.48\linewidth]{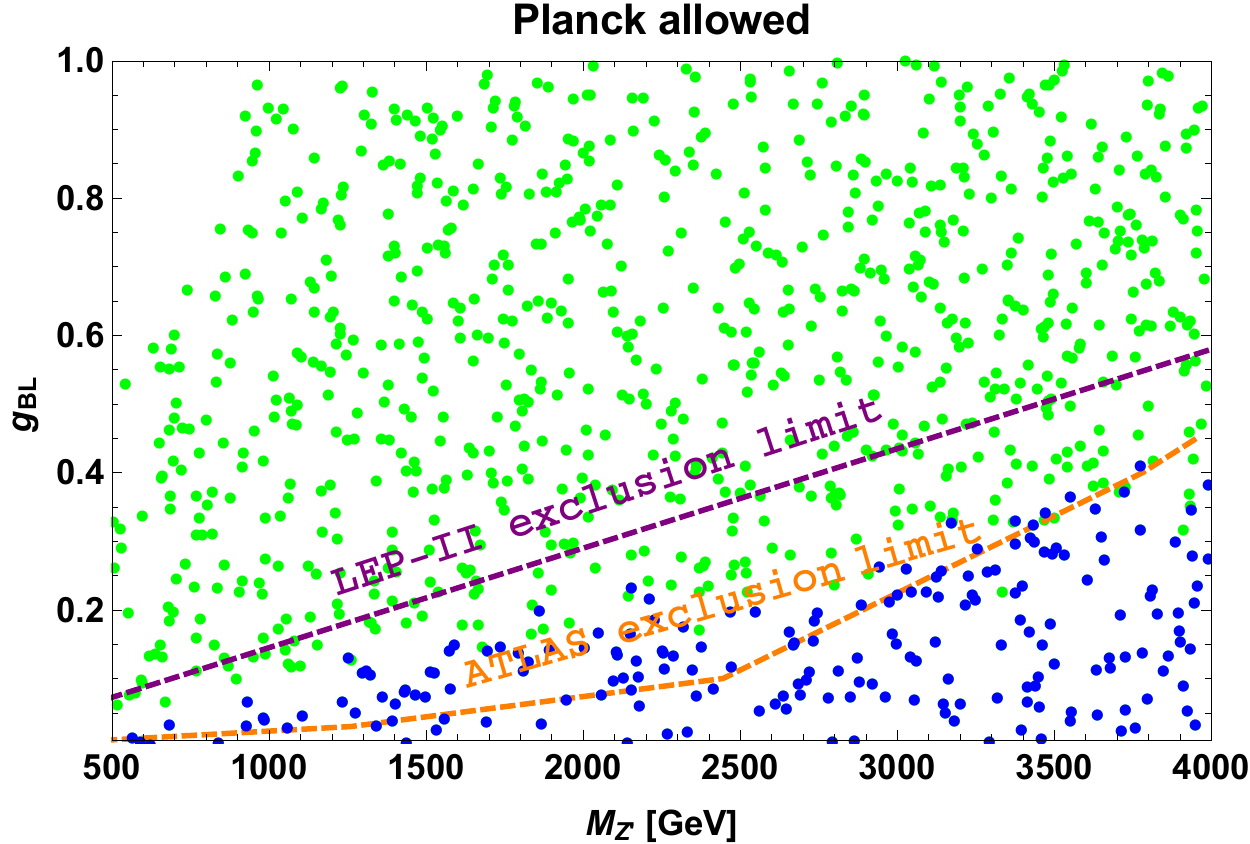}
\vspace{0.2 cm}
\includegraphics[width=0.48\linewidth]{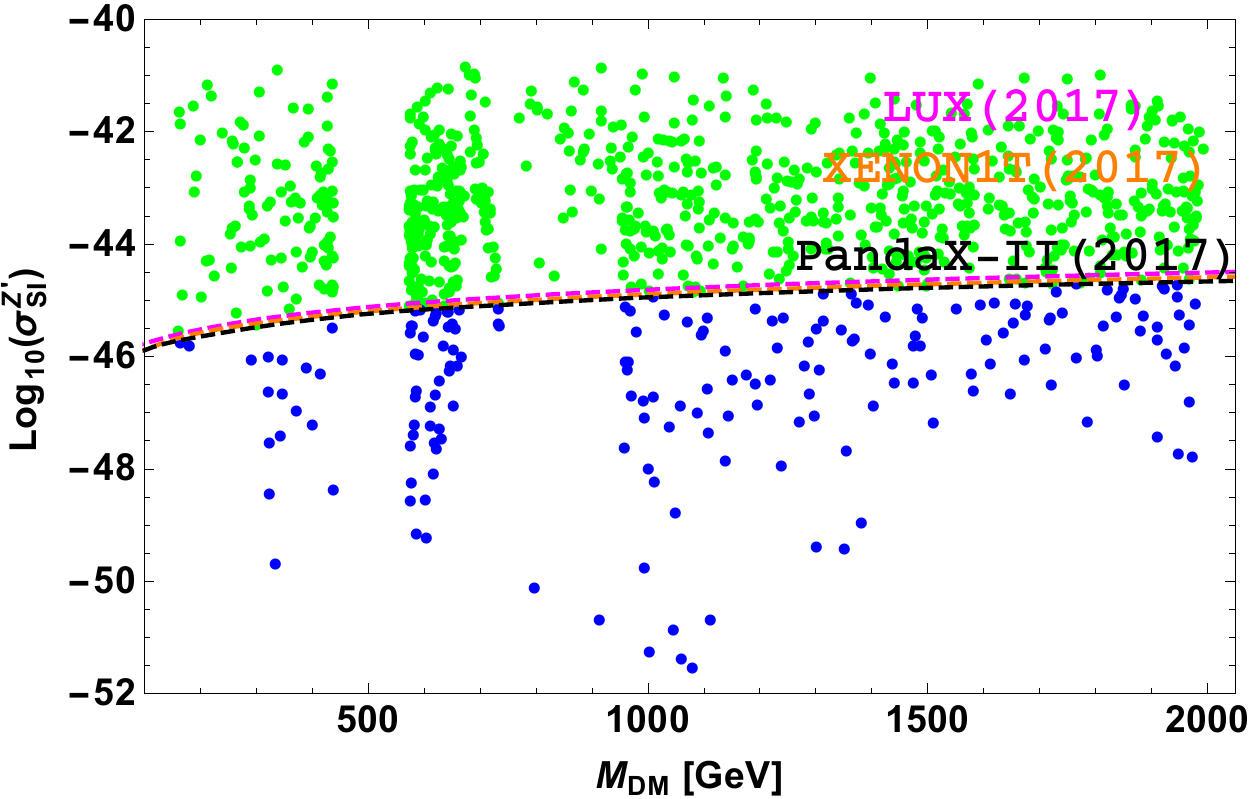}
\caption{Left panel depicts the $M_{Z^\prime}-g_{\rm BL}$ parameter space consistent with $3\sigma$ region of Planck relic density limit. Dashed lines represent the ATLAS \cite{TheATLAScollaboration:2015jgi} and LEP-II \cite{Schael:2013ita} limits. Right panels shows the SI WIMP-nucleon cross section for the parameter space shown in the left panel. Dashed lines represent the recent bounds dictated by PandaX-II \cite{Cui:2017nnn}, XENON1T \cite{Aprile:2017iyp} and LUX \cite{Akerib:2016vxi}.}\label{DD}
\end{center}
\end{figure}

For the parameter scan, we vary the DM mass $M_{\rm DM}$ from 50 GeV to 2 TeV, $M_{Z'}$ in the range 0.5 to 4 TeV, gauge coupling $g_{\rm BL}$ between 0 to 1, DM-scalar coupling in range $0.05$ to $0.1$ and the scalar masses $M_{H_1} = M_{H_2} = 1$ TeV. The left panel Fig. \ref{DD} shows the $M_{Z^\prime}-g_{\rm BL}$ parameter space, consistent with $3\sigma$ range of Planck limit on relic density. LEP-II and ATLAS exclusion bounds are denoted with magenta and orange dashed lines respectively. Here, the green data points violate the stringent upper limit on WIMP-nucleon SI cross section set by PandaX-II \cite{Cui:2017nnn} (visible from right panel). Therefore, the viable region of gauge parameters that survives all the experimental limits is the blue data points below ATLAS exclusion limit in the left panel. There is gap on the either side of $M_{\rm DM} = 500$ GeV, where no data point satisfies Planck data. This region corresponds to resonance in the $H_{1}$-mediated s-channel contribution. Though we are varying both $g_{\rm BL}$ and $M_{Z^\prime}$, the maximum gauge coupling is also insufficient for the $Z^\prime$ resonance to meet Planck limit. In the DM mass region of $750 - 900$ GeV, we notice few data points (green), that satisfy Planck relic density but inconsistent with direct detection experiments. This gap corresponds to the region starting just after $H_1$ resonance to till the mass range where $H_1 H_1$, $H_2H_2$ channel contribute to relic density (green curve in right panel of Fig. \ref{relicv2}). In this mass regime, with ($M_{Z^\prime} \sim 1.6 - 1.8$ TeV) and large values for $g_{\rm BL}$ can satisfy Planck relic density ($Z^\prime$ resonance). However, these large couplings provide large WIMP-nucleon cross section that violate direct detection upper limits.

Such $B-L$ scenarios with heavy scalars and fermions can be looked up in the light of LHC. The production of heavier higgs from gluon-gluon fusion can subsequently decay to various final states $pp(gg) \to H_{1,2} \to WW,ZZ,\gamma\gamma,\tau\tau,t\bar{t}$, have been investigated at CMS and ATLAS \cite{Khachatryan:2015cwa,CMS:2017vpy,Accomando:2016sge}. There are still other processes such as $pp \to Z^\prime, H_{1,2} \to N_i N_j$ need to be explored  if kinematically feasible in collider\cite{Accomando:2016sge}.
\subsection{Comment on indirect signals}
Fermi Large Area Telescope (LAT) and ground based MAGIC telescope \cite{Ahnen:2016qkx} have put constraints on DM annihilation rate to final states like $\mu^+\mu^-,\tau^+\tau^-, W^+W^-,b\overline{b},$ by measuring the gamma ray flux produced from them. These bounds are levied by considering $100\%$ annihilation of dark matter to particular final state particles to give Planck satellite consistent relic abundance. However, the present model deviates from such assumption as it opens up several DM annihilation channels such as fermion-anti fermion pair, SM gauge bosons, Higgs bosons in the final state in scalar and gauge portals, collectively contributing to total relic density. Further, gauge annihilation rate today is velocity suppressed and hence no signals are expected via $Z^\prime$ portal \cite{Rodejohann:2015lca,Berlin:2014tja}. The bounds from Fermi-LAT and MAGIC can  constrain scalar couplings to DM particle, which can also get restricted from spin-independent WIMP-nucleon cross section (scalar mediated) as well. However, in our analysis we give emphasis to gauge interactions rather than scalar interactions. 

\section{Realization of Leptogenesis in the present framework}
So far in the current framework, we have discussed DM phenomenology and  tree level neutrino mass. Now, one can also explain leptogenesis with the five dimension effective interaction of Dirac type with scalar singlet ($\phi_3$) in Eq.\eqref{dirac}. This induces the decay of lightest exotic fermion to SM Higgs and lepton in the final state after the breaking of ${ B-L}$ symmetry at TeV scale. Provided with the conversion relation $Y_B=\frac{8 N_F + 4 N_H}{22 N_F +13 N_H} Y_{ B-L}=\frac{28}{79} Y_{ B-L}$ \cite{Harvey:1990qw}, baryon asymmetry is produced through sphaleron transition, with $N_F$ and $N_H$ denote the number of fermion generations and Higgs doublets respectively.
\begin{figure}[h!]
\begin{center}
\includegraphics[width=0.3\linewidth]{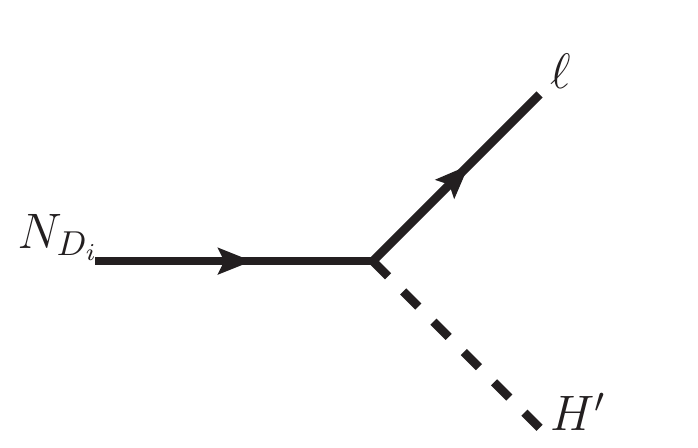}
\includegraphics[width=0.35\linewidth]{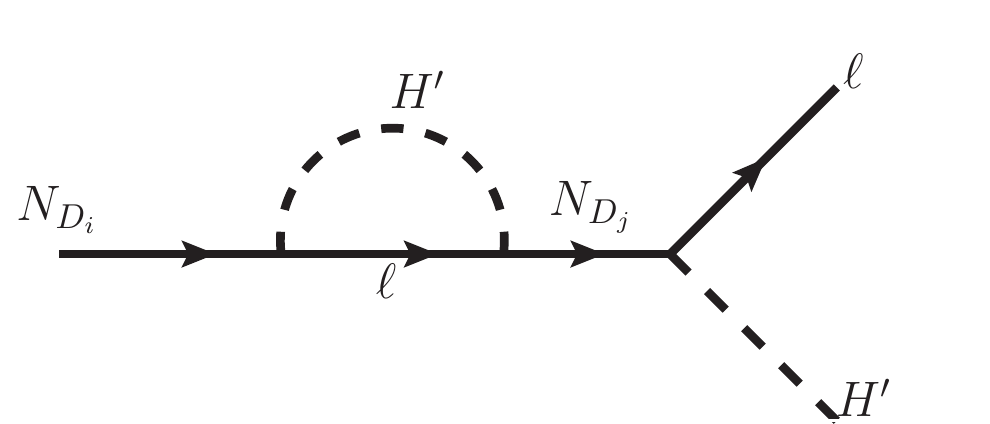}
\includegraphics[width=0.35\linewidth]{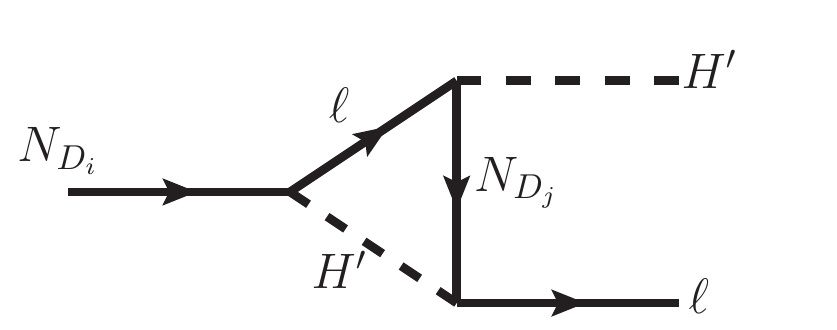}
\caption{Tree and loop level decay of lightest right-handed fermion.}
\label{rhndecay}
\end{center}
\end{figure}
As discussed earlier in section IIIB, we have a mass spectrum of heavy neutrinos i.e.,  $M_{1}\sim M_{2} < M_{3} \sim M_{4}$. Considering this case, the asymmetry can be generated from the decay of the lightest mass eigenstates $N_{D_1}$. The tree and loop level Feynman diagrams are shown in Fig. \ref{rhndecay} and the general expression for the CP asymmetry  is given by
\begin{equation}
\epsilon_i =\frac{1}{3}\times\frac{ \Gamma (N_{D_i}\rightarrow \ell   H^\prime)-\Gamma (N_{D_i}\rightarrow \overline{\ell}    \overline{{H^\prime}})}{ \Gamma (N_{D_i}\rightarrow \ell        H^\prime)+\Gamma (N_{D_i}\rightarrow \overline{\ell}    \overline{H^\prime})}\,,
\end{equation}
Overall factor arises as the decay mode violates $B-L$ by $1/3$ units. Interference of tree level decay  with one loop self energy and vertex correction gives non-zero CP asymmetry, can be written as
\begin{eqnarray}
\epsilon_i=\frac{1}{3}\times\frac{M_i}{M_j}\frac{\Gamma_j}{M_j} \left(\frac{V}{2}+S\right)\frac{{\rm Im} \left((\tilde{Y^\prime} {\tilde{Y^\prime}}^\dagger)^2_{ij}\right)}{(\tilde{Y^\prime} {\tilde{Y^\prime}}^\dagger)_{ii}(\tilde{Y^\prime} {\tilde{Y^\prime}}^\dagger)_{jj}}\,.
\end{eqnarray}
Here, the modified Yukawa coupling matrix $\tilde{Y^\prime}$ is given by
\begin{equation}
\tilde{Y^\prime}=Y^\prime U_N,\hspace{3mm}  Y^\prime =\frac{v_3}{\sqrt{2} \Lambda}\begin{pmatrix}
0 & Y^\prime_{12} & Y^\prime_{13} & Y^\prime_{14} \\
0 & Y^\prime_{22} & Y^\prime_{23} & Y^\prime_{24}\\
0 & Y^\prime_{32} & Y^\prime_{33} & Y^\prime_{34}
\end{pmatrix}, 
 \end{equation}
 $V$ and $S$ denote the vertex and self-energy contributions respectively, given by  
\begin{eqnarray}
&& V=2 \frac{{M_j}^2}{{M_i}^2}\left[\left(1+ \frac{{M_j}^2}{{M_i}^2}\right){\rm log}\left(1+ \frac{{M_j}^2}{{M_i}^2}\right)-1\right]\,,\\
&& S=\frac{{M_j}^2 \Delta {M_{ij}}^2}{(\Delta {M_{ij}}^2)^2+{M_i}^2 {\Gamma_j}^2},\hspace{3mm} \frac{\Gamma_j}{M_j}=\frac{(\tilde{Y^\prime} {\tilde{Y^\prime}}^\dagger)_{jj}}{8\pi},\hspace{4mm} \Delta {M_{ij}}^2={M_j}^2-{M_i}^2.
\label{selfv}
\end{eqnarray}
In the above expression $\Gamma_j$ is the tree level decay width of the corresponding heavy fermion. Leptogenesis from the decay of heavy Majorana neutrinos with a hierarchical mass spectrum has been widely discussed in the literature \cite{Dev:2017trv,Pascoli:2006ie,Abada:2006ea}. These studies mainly focus different cases like single flavor approximation and flavor consideration. With one flavor approximation, Casas-Ibarra bound on right-handed neutrino mass is of the order $ \mathcal O(10^9)$ GeV \cite{Davidson:2002qv}, to explain the observed baryon asymmetry. However, we opt for resonance enhancement of CP asymmetry in case of quasi degenerate Majorana neutrinos with a mass scale as low as TeV \cite{Pilaftsis:2003gt,Asaka:2018hyk}.\\  
%With a hierarchical mass structure of right-handed fermions, both self energy and vertex diagrams equally contribute to the CP asymmetry. And the Ibara bound on the lightest right-handed neutrino mass ($\gg 10^9$ GeV) from neutrino oscillation data, in order to explain the observed baryon asymmetry, has already been discussed in the literature  \cite{Davidson:2002qv}.
\begin{figure}[h!]
\begin{center}
\includegraphics[width=0.18\linewidth]{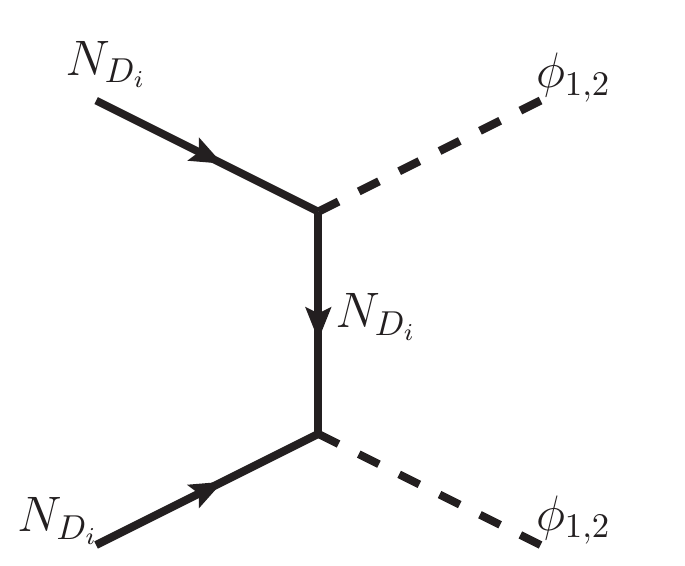}
\includegraphics[width=0.28\linewidth]{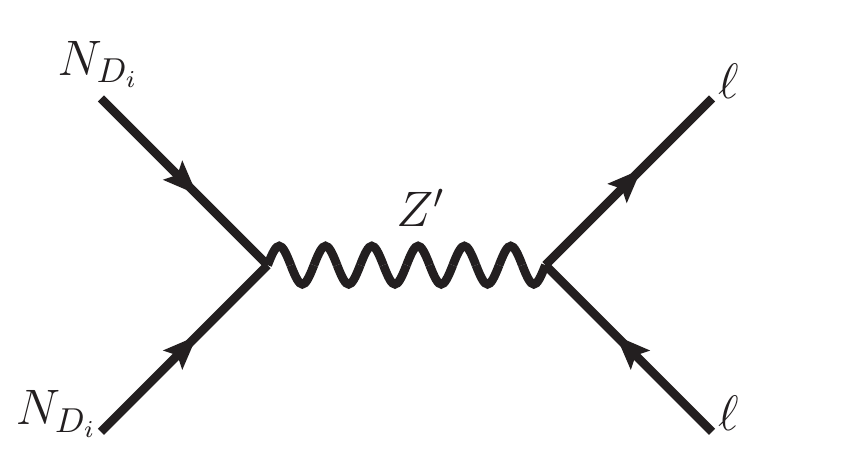}
\includegraphics[width=0.28\linewidth]{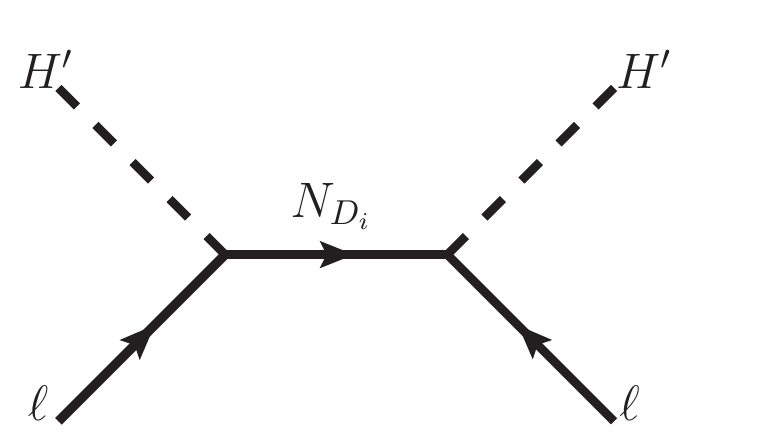}
\includegraphics[width=34mm,height=26mm]{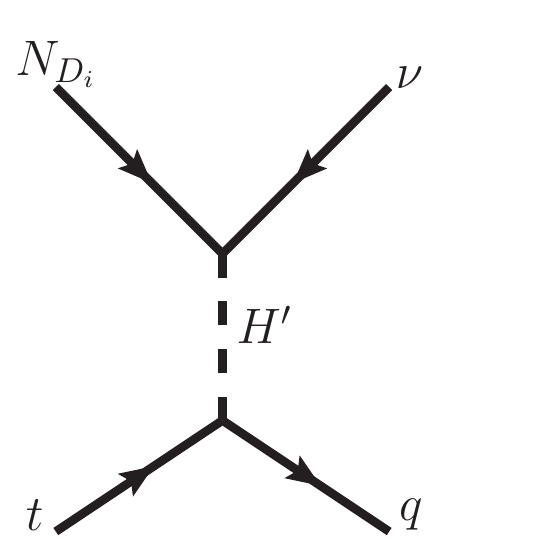}
\caption{Scattering procesess that effect the yield of right-handed fermion and lepton.}\label{rhnsctrng}
\end{center}
\end{figure}
Following \cite{Pilaftsis:2003gt}, we consider resonant enhancement in CP asymmetry with the fermion mass splitting $\Delta {M_{ij}}^2 \approx M_i \Gamma_j $. Thus Eq.\eqref{selfv}\, gives maximum contribution from self energy with $S \approx \frac{M_j}{\Gamma_j} \gg 1$ and the vertex contribution can be safely neglected. Thus CP asymmetry can be reduced to the form 
\begin{equation}
\epsilon_i \approx \frac{1}{3}\times\frac{1}{2}\frac{ {\rm Im}\left[\left(\tilde{Y^\prime}{\tilde{Y^\prime}}^\dagger\right)^2_{ij}\right]}{\left(\tilde{Y^\prime}{\tilde{Y^\prime}}^\dagger\right)_{ii}\left(\tilde{Y^\prime}{\tilde{Y^\prime}}^\dagger\right)_{jj}}\,,
\end{equation}
Hence from the above expression, considering the Yukawa couplings in similar order, the CP parameter can be enhanced to achieve a value of order $1$.
%==================================================
\subsection{Boltzmann Equations}
%===============================================
The final baryon asymmetry depends on the efficiency of leptogenesis, which could be derived from the dynamics of relevant Boltzmann equations. When the gauge interaction rate is more than the Hubble expansion, particles attain thermal equilibrium and are subjected to the chemical equilibrium constraints. Hence the Boltzmann equations are so important to study the particle number density after the chemical or kinetic decoupling from the thermal bath in a specific temperature regime. Lepton number violation demands the decay of the heavy fermion to be out of equilibrium to satisfy the Sakharov's condition. The Boltzmann equations for the evolution of the number densities of right-handed fermion and lepton, written in terms of yield parameter (ratio of number density to entropy density) are given by \cite{Plumacher:1996kc, Iso:2010mv}
\begin{eqnarray}
&& \frac{d Y_{N}}{dz}=-\frac{z}{s H(M_1)} \left[\left( \frac{Y_N}{{Y^{eq}_N}}-1\right)\gamma_D+\left( \left(\frac{{Y_N}}{{Y^{eq}_N}}\right)^2-1\right)(\gamma_{Z'}+\gamma_{\phi})\right],\\
&& \frac{d Y_{ B-L}}{d z}= -\frac{z}{s H(M_1)} \left[ \epsilon_1 \left( \frac{Y_N}{{Y^{eq}_N}}-1\right)\gamma_D+\left(\frac{\gamma_D}{2}+\gamma_w\right) \frac{Y_{\rm B-L}}{{Y^{eq}_{\ell}}} \right],
\end{eqnarray}
where $H,s$ represent the Hubble rate and entropy density, $z = M_1/T$ and the equilibrium number densities are given by
\begin{eqnarray}
%H(T)=\frac{4 {\pi}^3 g_\star}{45} \frac{T^2}{M_{\rm pl}},\hspace{3mm} %\hspace{3mm}  \text{where,} \hspace{3mm}    M_{\rm pl}=1.22\times 10^{19} ~\text{GeV},\\
Y^{eq}_N= \frac{135 \zeta{(3)} g_N}{16 {\pi}^4 g_\star} z^2 K_2(z), \hspace{3mm} {Y^{eq}_\ell}= \frac{3}{4} \frac{45 \zeta(3) g_\ell}{2 {\pi}^4 g_{\star}}\,.
\end{eqnarray}
Here, $K_{1,2}$ denote modified Bessel functions, $g_\star=106.75$ (total number of relativistic degrees of freedom), $g_\ell=2$ and $g_N=2$ denote the degrees of freedom of lepton and right-handed fermions respectively. The decay rate $\gamma_D$ is given by
\begin{equation}
\gamma_D = s Y_{N}^{eq}\Gamma_1 \frac{K_1(z)}{K_2(z)}.
\end{equation}
The scattering processes that can change the number density of the right-handed neutrino are depicted in the first two panels of Fig. \ref{rhnsctrng} i.e. $\gamma_{\phi}$ (pair of scalars in the final state) and $\gamma_{Z'}$ (fermion-antifermion pair in final state via $Z^\prime$). Third and fourth panels stand for the washout processes ($\gamma_w$) that reduce the lepton asymmetry. The details for computing the reaction rates are provided in \cite{Plumacher:1996kc,Iso:2010mv}. 

\begin{figure}[t!]
\begin{center}
\includegraphics[width=0.48\linewidth]{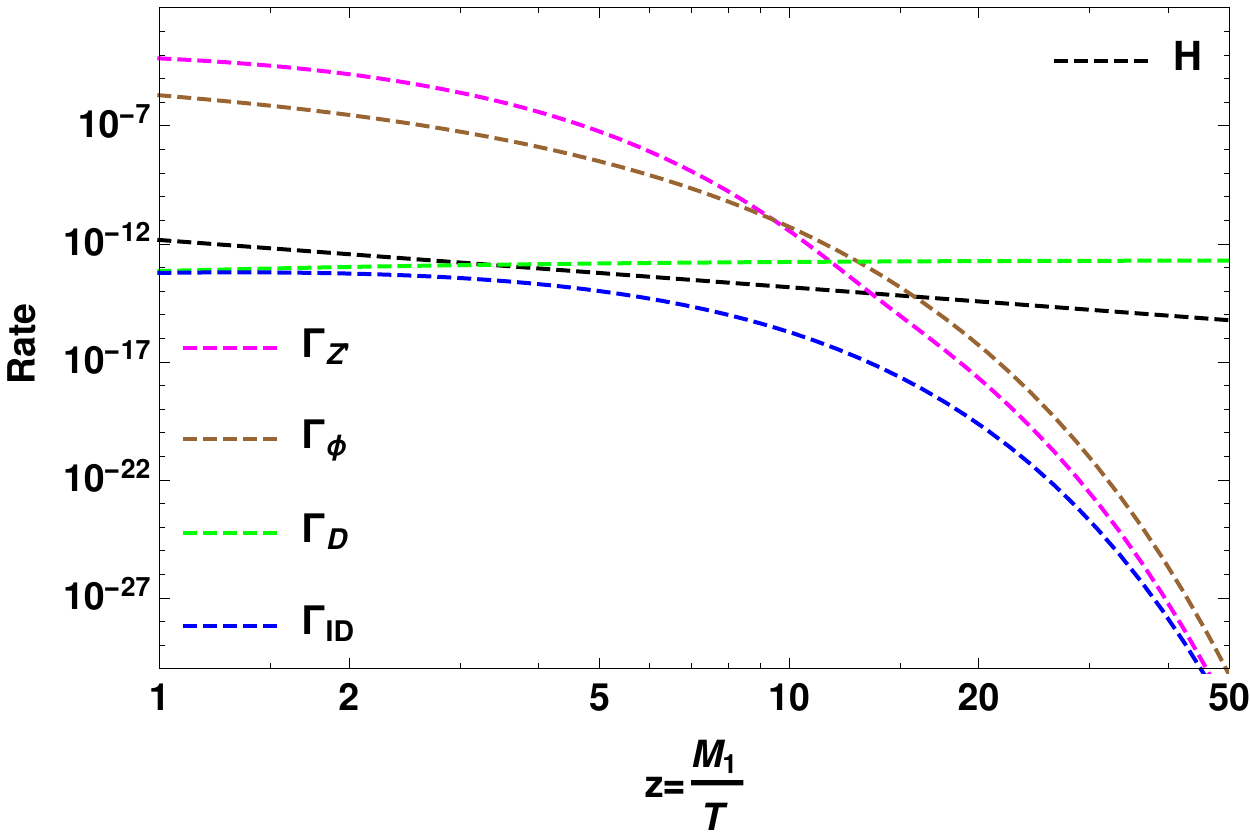}
\includegraphics[width=0.48\linewidth]{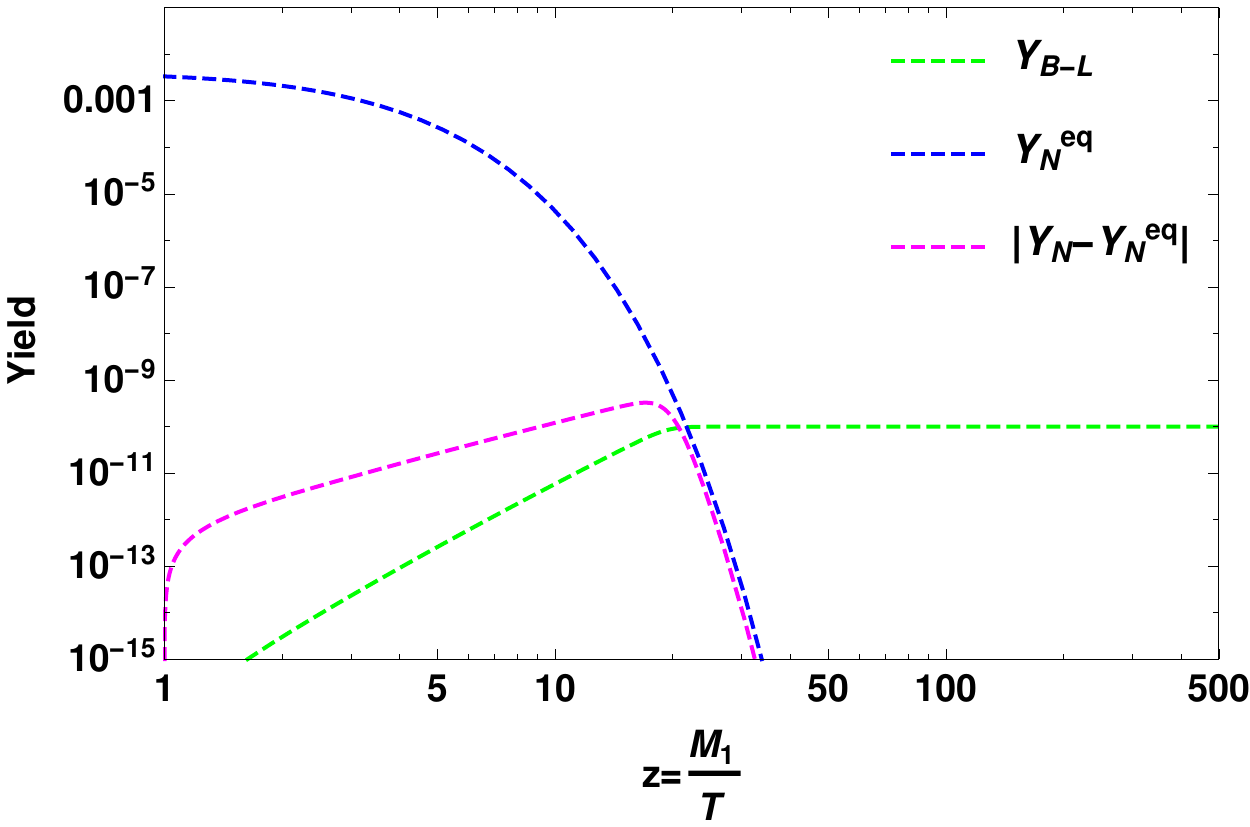}
\caption{Left panel represents the reaction rates of decay, inverse decay and scattering process and right panel shows the evolution of right-handed fermion and $B-L$ yield, obtained for $M_1 = 1$ TeV.}\label{beqn}
\end{center}
\end{figure}

We project the reaction rates of decay ($\Gamma_{\rm D}$), inverse decay ($\Gamma_{\rm ID}$) and relevant scattering processes in the left panel of Fig. \ref{beqn}. Both $\Gamma_{Z'}$ ($=\gamma_{Z'}/(sY_N^{eq})$) and $\Gamma_{\phi}$ ($=\gamma_{\phi}/(sY_N^{eq})$) play a significant role in reducing the heavy fermion number density. In the analysis, we took Yukawa coupling $\left(\tilde{Y^\prime_{ij}}\right)$ of order $10^{-7}$ for decay and inverse decay, $M_{Z'}=3$ TeV and $g_{\rm BL}=0.1$ (for $\Gamma_{Z'}$) and Majorana coupling to be order $0.1$ (for $\Gamma_{\phi}$).
Right panel of Fig. \ref{beqn} represents the evolution of right-handed fermion and $B-L$ yield. The scatterings make $Y_N$ stay close to thermal equilibrium. The obtained $B-L$ asymmetry is of the order $10^{-10}$ for a Yukawa coupling of order $\approx 10^{-7}$ and CP violation parameter $\epsilon_1=0.02$. Thus, the value of baryon asymmetry can be computed using the relation $Y_B = \frac{28}{79} Y_{B-L}$.  
\subsection{A note on flavor effects}
\begin{figure}[t!]
\begin{center}
\includegraphics[width=0.48\linewidth]{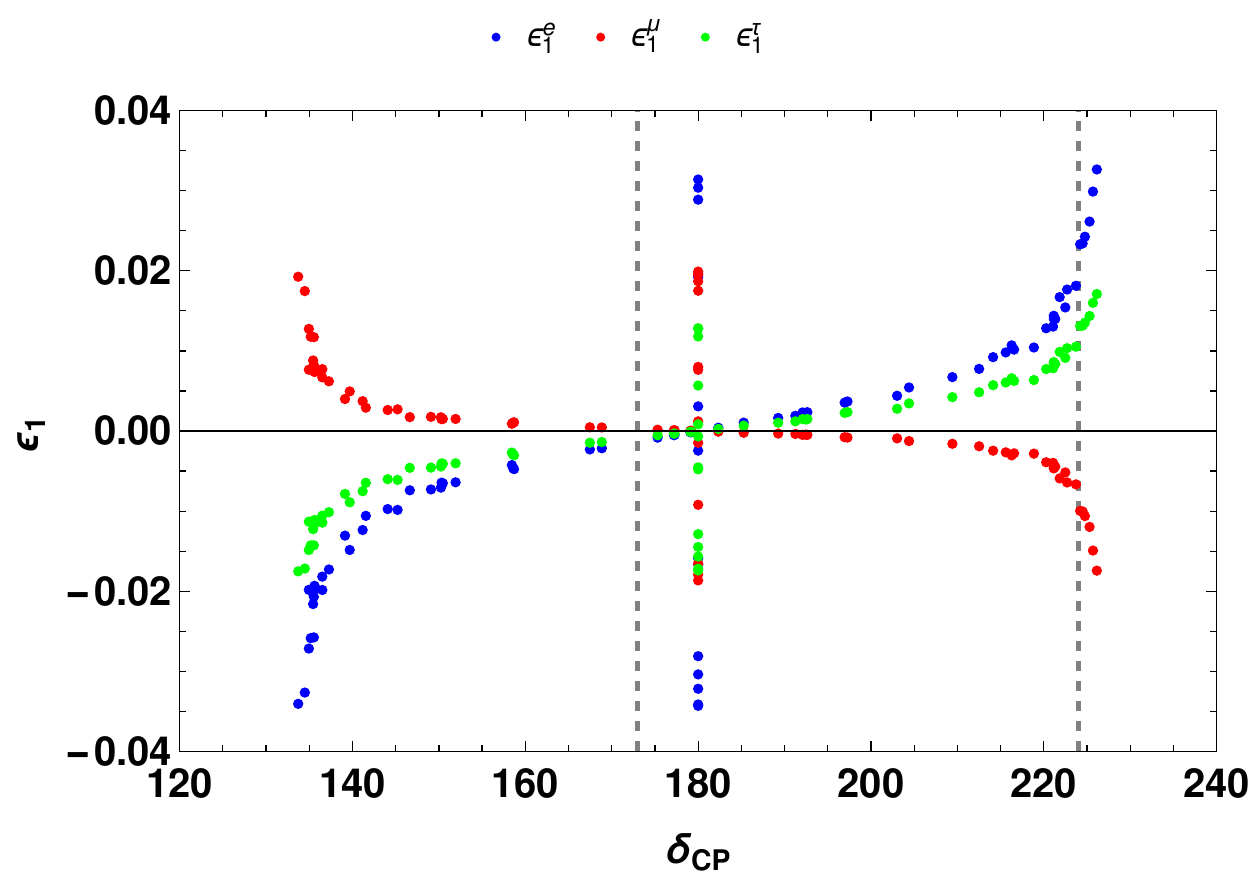}
\caption{Dependence of CP asymmetry on Dirac CP phase. Vertical dashed lines correspond to $1\sigma$ allowed region of global fit data \cite{Esteban:2020cvm}.}\label{CPV}
\end{center}
\end{figure}

The generic case of one flavor approximation is probable in the high temperature regime ($T>10^{12}$ GeV), where all the interactions mediated by Yukawa couplings are out of equilibrium. But temperatures below $10^{12}$ GeV provoke various charged lepton Yukawa couplings to come to equilibrium and then the flavor effects play a vital role in generating the final lepton asymmetry. Since in a temperature regime below $10^5$ GeV, all the Yukawa interactions are in equilibrium, the asymmetry is stored in the individual lepton sector. The detailed discussion of flavor effects on the lepton asymmetry generated from the decay of right-handed neutrinos has been explored in literature \cite{Pascoli:2006ci,Antusch:2006cw,Nardi:2006fx,Abada:2006ea,Granelli:2020ysj,Dev:2017trv}. Consideration of flavor effects relaxes the lower bound on heavy Majorana masses and therefore provides the flexibility to bring down the scale of leptogenesis  \cite{Abada:2006fw}. Since flavor effects are important in low scale leptogenesis, we discuss briefly their impact in the present framework. To start with, we compute the CP asymmetry in the resonant condition for individual lepton flavor $\alpha$ as follows \cite{Dev:2015cxa}
\begin{equation}
\epsilon^\alpha_{i}\approx \frac{1}{6} \times \sum_{j,i \neq j} \left[ \frac{\text{Im}\left[(\tilde{Y^\prime} {\tilde{Y^\prime}}^{\dagger})_{ij} \tilde{Y^\prime_{\alpha i}}^* \tilde{Y^\prime_{\alpha j}} \right]}{(\tilde{Y^\prime} \tilde{{Y^\prime}^{\dagger}})_{ii} (\tilde{Y^\prime} \tilde{{Y^\prime}^{\dagger}})_{jj}}\right]\,.
\end{equation}
Fig. \ref{CPV} projects the dependence of CP asymmetry in individual lepton flavors on the Dirac CP phase. 
The total baryon asymmetry depends on the washout parameter given by
\begin{eqnarray}
K_\alpha = \frac{ \Gamma_{N_{D_1}\rightarrow \ell_{\alpha} H^\prime}+\Gamma_{N_{D_1}\rightarrow \overline{\ell_{\alpha}} \overline{H^\prime}}}{H(M_1)}=\frac{\tilde{m_\alpha}}{m_{\star}}, \quad K = \sum_\alpha K_\alpha = \frac{\tilde{m}}{m_{\star}}. 
\end{eqnarray}
Here $m_\star \approx 3\times 10^{-3}$ eV is the equilibrium neutrino mass and $\tilde{m}$ is the effective neutrino mass, given by
\begin{eqnarray}
\tilde{m} = \sum_\alpha \tilde{m_\alpha}, \quad \tilde{m_\alpha}=\frac{(\tilde{Y^\prime_{\alpha 1}}^* \tilde{Y^\prime_{\alpha 1}}) v^2}{M_1}.
\end{eqnarray}
%The efficiency factor for different flavors is given by \cite{Pascoli:2006ci,Abada:2006fw}
%\begin{eqnarray}
%\eta(\tilde {m_l})=\left[\left(\frac{\tilde{m_l}}{8.25 \times 10^{-3}~{\rm eV}}\right)^{-1}+\left(\frac{0.2 \times 10^{-3} ~{\rm eV}}{\tilde{m_l}}\right)^{-1.16}\right]^{-1}\;,.
%\end{eqnarray}
The flavored Boltzmann equation for generating the lepton asymmetry is given by \cite{Antusch:2006cw}
\begin{eqnarray}
\frac{d Y^{\alpha}_{ B-L_\alpha}}{d z}= -\frac{z}{s H(M_1)} \left[ \left( \epsilon^\alpha_1 \gamma_D \left( \frac{Y_N}{{Y^{eq}_N}}-1\right)\right)-\left(\frac{\gamma^{\alpha}_D}{2}+\gamma^\alpha_w\right)\frac{A_{\alpha \alpha}Y^\alpha_{\rm B-L_\alpha}}{{Y^{eq}_{\ell}}}\right].
\end{eqnarray}
Here,
\begin{equation}
\gamma_D^\alpha = s Y_{N}^{eq}\Gamma_1^\alpha \frac{K_1(z)}{K_2(z)}, \quad \gamma_D = \sum_\alpha \gamma^\alpha_D,\nn\\
\end{equation}
\begin{equation}
A=\begin{pmatrix}
-\frac{151}{179} && \frac{20}{179} && \frac{20}{179}\\
\frac{25}{358} && -\frac{344}{537} && \frac{14}{537}\\
\frac{25}{358} && \frac{14}{537}  && -\frac{344}{537} \\
\end{pmatrix}.\nn\\
\end{equation}
%Thus the total baryon asymmetry can be expressed as following
%\begin{equation}
%Y_B = \left(-\frac{28}{79}\right) Y_L=\frac{-28/79}{g_\star}\left[\epsilon_e \eta\left(\frac{151}{179}\tilde{m_e}\right)+\epsilon_\mu \eta\left(\frac{344}{537} \tilde{m_\mu}\right)
%+\epsilon_\tau \eta\left(\frac{344}{537}\tilde{m_\tau}\right)\right]\;.
%\end{equation}
\begin{figure}[t!]
\begin{center}
\includegraphics[width=0.48\linewidth]{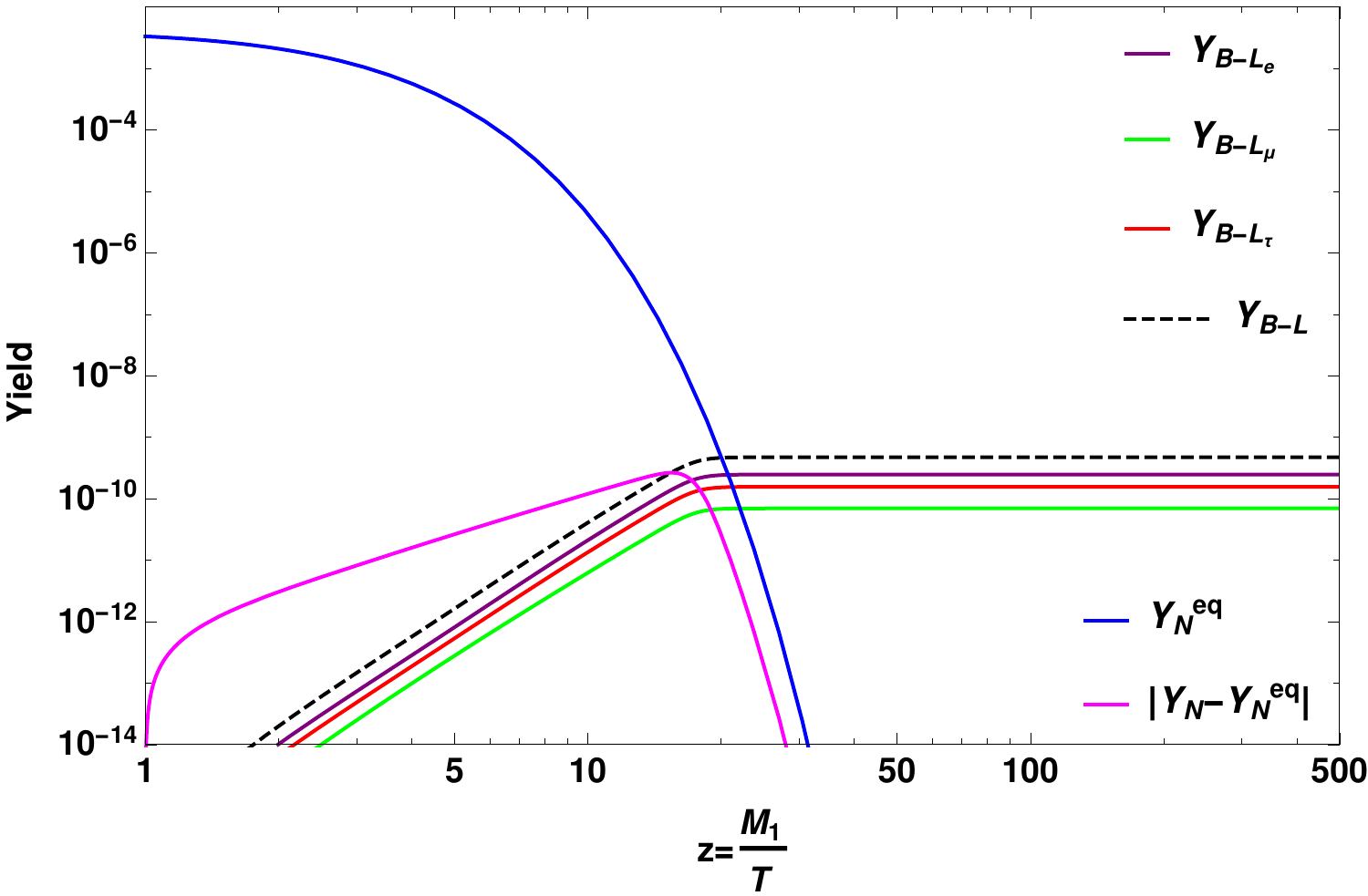}
\includegraphics[width=0.48\linewidth]{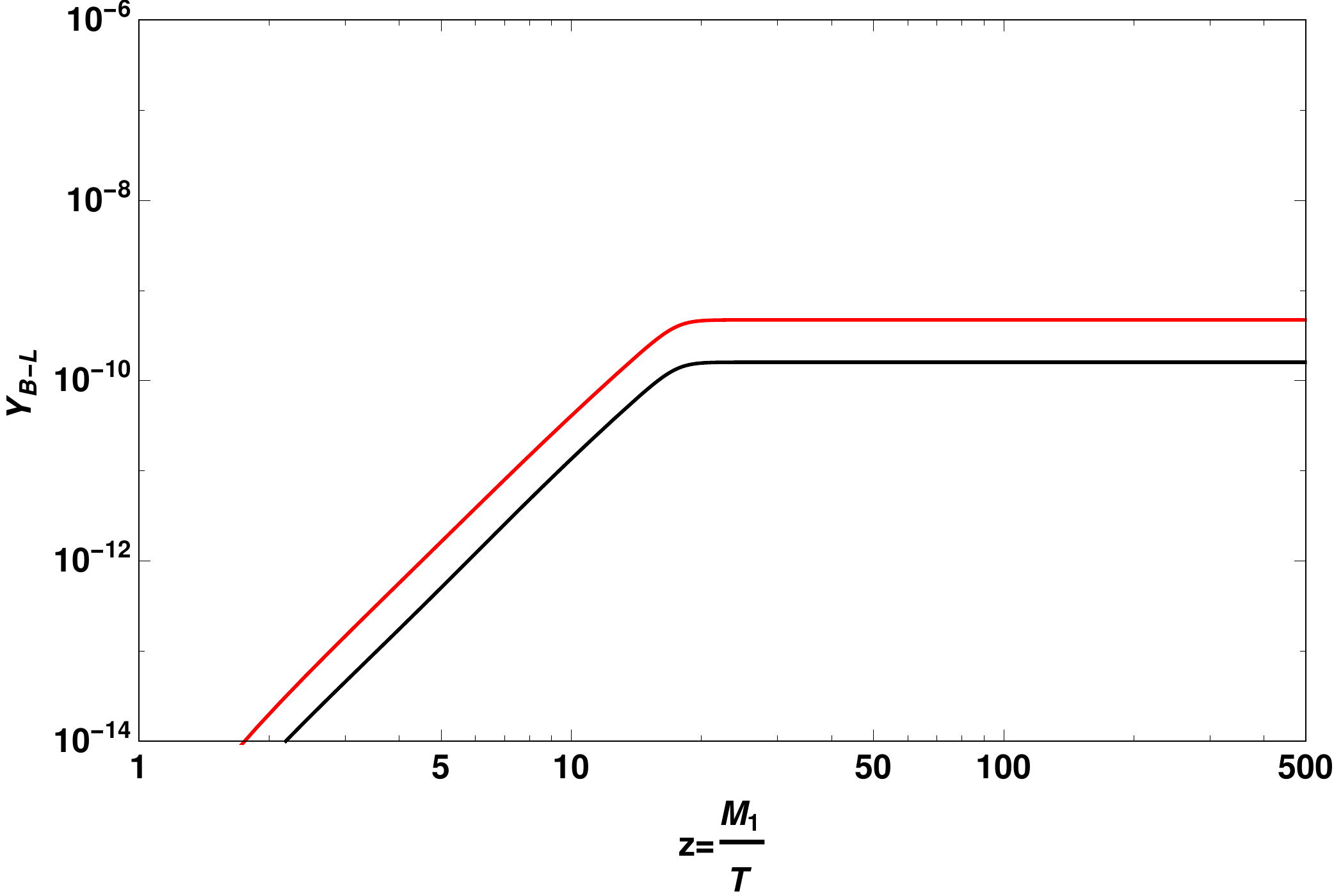}
\caption{Left panel projects $B-L$ yield with the inclusion of flavor effects. Right panel projects the enhancement in $Y_{B-L}$ due to three-flavor calculation (red curve) over one-flavor approximation (black curve).}\label{beqn2}
\end{center}
\end{figure}

Left panel of Figure. \ref{beqn2} projects the evolution of asymmetries in individual lepton flavors. We choose a specific benchmark from Fig. \ref{CPV}, for $\delta_{\rm CP} = 219^\circ$, we have $\epsilon_e = 10^{-2}$, $\epsilon_{\mu} = -3\times10^{-3}$, $\epsilon_{\tau} =6\times 10^{-3}$ and $K_e = 9\times 10^{-1}$, $K_\mu = K_\tau = 2.5\times 10^{-2}$. We notice a slight enhancement in $B-L$ asymmetry due to flavor effects as projected in the right panel. But this enhancement can be more prominent in the strong washout region ($K_\alpha>1$). Since in this regime, the decay of heavy fermion to a specific final state lepton flavor can be washed out by the inverse decay of any flavor with one flavor approximation unlike the flavored case \cite{Abada:2006ea}. 

%\red{However, a detailed analysis for the connection between flavor effects in Leptogenesis and neutrino oscillation parameters is beyond the scope of this work and is well explored in the literature \cite{Pascoli:2006ci}.}
%\begin{table}[tb!]
%\begin{center}
%%================================================
%\begin{tabular}{|c|c|c|c|c|c|c|}
%	\hline
%		$M_1$ & $Y_{ij}$	& $\epsilon_{e}$	& $\epsilon_{\mu}$ & $\epsilon_{\tau}$ &  $Y_B$\\
%	\hline
%	$\mathcal{O}$ (TeV) & $\mathcal{O}(10^{-6})$ & $0.003$ & $0.002$ & $0.008$ & $\mathcal{O}(10^{-10})$ \\
%	\hline
%\end{tabular}
%\caption{Sample benchmark.}
%\label{model_charges}
%\end{center}
%\end{table}
%=====================================================
\section{Constraints on new gauge parameters from quark and lepton sectors}
\begin{figure}[htb]
\centering
\includegraphics[width=0.3\linewidth]{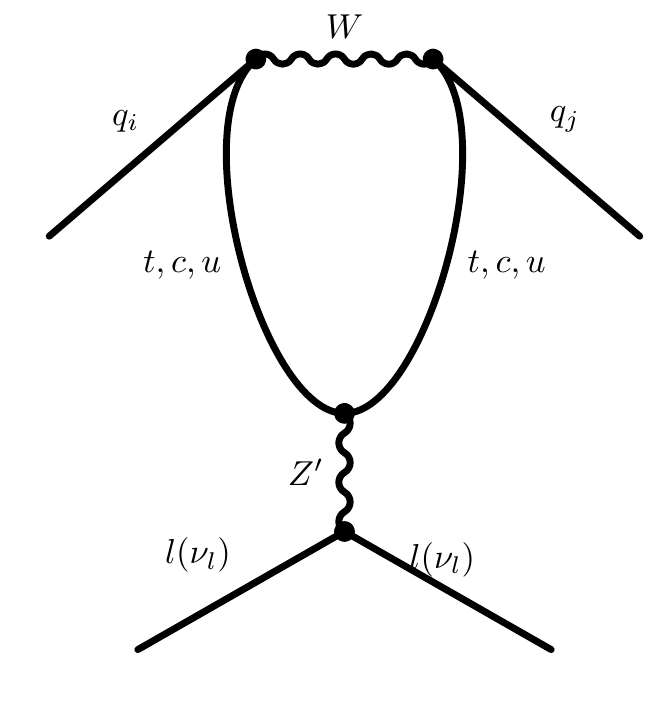}
\caption{One loop penguin diagram of $q_i \to q_j l\bar l (\nu_l \bar \nu_l)$ mediated by $Z^\prime$ gauge boson.}\label{Fig:penguin}
\end{figure}

%======================================================
Since the $Z^\prime q_i q_j$ interaction term is not allowed in the proposed model, the  leptonic/semileptonic $B, D, K$ modes involving the quark level transitions   $q_i \to q_j ll(\nu_l \bar \nu_l)$ ($q_i=b,c, s$, $q_j=u,d,s$ and  $l$ is any charged lepton) can only occur at one loop level via $Z^\prime$ boson as shown in Fig. \ref{Fig:penguin}\,. 
We  mainly focus on the existing data on the branching ratios of  $B$ meson channels to constrain the $M_{Z^\prime}-g_{\rm BL}$ plane.  The Lepton Flavor Violating (LFV) decay modes of $B$ meson and $\tau(\mu)$ lepton are not allowed due to the absence of $Z^\prime l_i l_j$ coupling, thus we use the branching ratio of only possible $\tau \to \mu \nu_\tau \bar \nu_\mu$ process for this purpose.  In SM, the explicit form of the effective  Hamiltonian which is responsible for leptonic/semileptonic $b \to q(=s,d) ll$ transitions is given by \cite{Beneke:2004dp, Buchalla:1993wq, Fajfer:2015mia, Fajfer:2012nr}
\bea
\mathcal{H}^{\rm eff}= -\frac{4G_F}{\sqrt{2}}V_{tb}V_{tq}^*\left (\sum_{i=1,\cdots 10,S,P} C_i \mathcal{O}_i+ \sum_{i=7,\cdots 10,S,P}C_i' \mathcal{O}_i'\right ), \label{SM-Ham-Flavor}
\eea 
where $G_F$ is the Fermi constant, $V_{qq^\prime}$ is the product of CKM matrix elements. 
Here $\mathcal{O}_i$'s are the effective operators,  defined as 
\bea
O_7^{(\prime)} &=&\frac{e}{16 \pi^2} \Big(\bar q \sigma_{\mu \nu}
(m_{q} P_{L(R)} + m_b P_{R(L)} ) b\Big) F^{\mu \nu}\,, \nn\\
O_9^{(\prime)}&=& \frac{\alpha_{\rm em}}{4 \pi} (\bar q \gamma^\mu P_{L(R)} b)(\bar l \gamma_\mu l)\;,~~~~~~~ O_{10}^{(\prime)}= \frac{\alpha_{\rm em}}{4 \pi} (\bar q \gamma^\mu 
P_{L(R)} b)(\bar l \gamma_\mu \gamma_5 l)\;,
\eea
with $\alpha_{\rm em}$ is the fine structure constant,   $P_{L,R}=(1\mp \gamma_5)/2$ are the projection operators and $C_i^{(\prime)}$'s are the corresponding Wilson coefficients.  The values of Wilson coefficients in the SM  are taken from \cite{Hou:2014dza, Buchalla:1993wq, Buchalla:1998ba, Misiak:1999yg}. The primed operators are absent in the SM, however the respective  coefficients  may be non-zero in the presence of $Z^\prime$ boson arising due to the $\rm U(1)_{B-L}$ gauge extension. Using  the interaction terms of SM fermions with $Z^\prime$ from Eq.\eqref{Lagrangian}\,,  the effective Hamiltonian for $b \to q ll$ processes  is given by
\bea \label{ZP-Ham-Flavor}
\mathcal{H}^{Z^\prime}=\frac{G_F}{12 \sqrt{2} \pi^2}V_{tb}V_{tq}^* F\left(\frac{m_t^2}{M_W^2}\right) \frac{{g^2_{\rm BL}}}{M_{Z^\prime}^2}(\bar q \gamma^\mu P_{L} b)(\bar l \gamma_\mu l)\,,
\eea 
where $F\left(\frac{m_t^2}{M_W^2}\right)$ is  the loop function that is order one ($F\left(\frac{m_t^2}{M_W^2}\right) \approx 1$) by using $m_t$ and $M_W$ from PDG \cite{Tanabashi:2018oca}. 
Now comparing Eq.\eqref{ZP-Ham-Flavor} with \ref{SM-Ham-Flavor}\,, we obtain an additional Wilson coefficient  contribution to the SM as
\bea
C_9^{\rm NP}=-\frac{1}{12 \pi \alpha_{\rm em}} \frac{{g^2_{\rm BL}}}{M_{Z^\prime}^2}\,.
\eea
The effective Hamiltonian for rare decay  processes mediated by $b \to q(=d,s) \nu_l \bar \nu_l$ transitions are given by  \cite{Altmannshofer:2009ma}
\bea
\mathcal{H}_{\nu}^{\rm eff} = \frac{-4G_F}{\sqrt{2}} V_{tb}V_{tq}^*\left(C^\nu_L \mathcal{O}^\nu_L +C^\nu_R \mathcal{O}^\nu_R \right) + h.c., \label{nu-ham}
\eea
where the $\mathcal{O}^\nu_{L(R)}$ effective operators are defined as 
\bea
\mathcal{O}^\nu_L = \frac{\alpha_{\rm em}}{4 \pi} \left(\bar{q}\gamma_\mu P_L b \right) \left(\bar{\nu_l} \gamma^\mu \left(1-\gamma_5\right)\nu_l\right),
 \hspace{1cm} \mathcal{O}^\nu_R = \frac{\alpha_{\rm em}}{4 \pi} \left(\bar{q}\gamma_\mu P_R b \right) \left(\bar{\nu_l} \gamma^\mu \left(1-\gamma_5\right)\nu_l\right)\,.
\eea
Here the Wilson coefficient $C^\nu_{L}$ ($= -X\left(\frac{m_t^2}{M_W^2}\right)/\sin^2\theta_W$) is calculated by using the loop function $X \left(\frac{m_t^2}{M_W^2}\right)$ \cite{Misiak:1999yg, Buchalla:1998ba} and $C_R^\nu$ is negligible in the SM. The effective Hamiltonian in the presence of $Z^\prime$ is 
 \bea
 \mathcal{H}^{Z^\prime}_{\nu}=\frac{G_F}{24 \sqrt{2} \pi^2}V_{tb}V_{tq}^* F\left(\frac{m_t^2}{M_W^2}\right) \frac{{g^2_{\rm BL}}}{M_{Z^\prime}^2}\left(\bar{q}\gamma_\mu P_L b \right) \left(\bar{\nu_l} \gamma^\mu \left(1-\gamma_5\right)\nu_l\right),
 \eea
 which in comparison with Eq.\eqref{nu-ham} provides new contribution to $C_L$ Wilson coefficient as
 \bea
 C_{L}^{\nu \rm NP}=-\frac{1}{24 \pi \alpha_{\rm em}} \frac{{g^2_{\rm BL}}}{M_{Z^\prime}^2}\,.
 \eea
After collecting an idea on new Wilson coefficient contribution, we now proceed to constrain the new gauge parameters from the flavor observables, to be presented
in the subsequent subsections.

%===========================================
\subsection{$B \to (\pi, K) ll$}
%========================================
The branching ratio of $B \to K ll$ process with respect to $q^2$ is given by \cite{Bobeth:2007dw}
\bea
 \frac{d {\rm Br}}{d q^2 }= \tau_B \frac{G_F^2 \alpha_{\rm em}^2 |V_{tb}V_{ts}^*|^2}{2^8 \pi^5 M_B^3}\sqrt{\lambda(M_B^2, M_K^2, q^2)} \beta_l f_+^2(q^2) \Big ( a_l(q^2)+\frac{c_l(q^2)}{3} \Big )\;,
 \eea
 where,
 \bea
 a_l(q^2)&=&  q^2|F_P|^2+ \frac{\lambda(M_B^2, M_K^2, q^2)}{4}
 (|F_A|^2+|F_V|^2) \nn \\ &&+ 2 m_l (M_B^2-M_K^2+q^2) {\rm Re}(F_P F_A^*) +4 m_l^2 M_B^2 |F_A|^2 \;,\nn\\
  c_l(q^2)&=& -   \frac{\lambda(M_B^2, M_K^2, q^2)}{4} \beta_l^2 
 \left (|F_A|^2+|F_V|^2\right ),
 \eea
 with
\bea
F_V & =&\frac{2 m_b}{M_B} C_7^{\rm eff}+ C_9^{\rm eff} + C_9^{ \rm NP}, ~~~~~F_A = C_{10},\nn\\
F_P&=&  m_l C_{10} \Big[ \frac{M_B^2 -M_K^2}{q^2}\Big(\frac{f_0(q^2)}{f_+(q^2)}-1\Big) -1 \Big]\;,
\eea 
and  
\bea
\lambda(a,b,c) =a^2+b^2+c^2-2(ab+bc+ca),~~~~~\beta_l = \sqrt{1-4 m_l^2/q^2}\;.
\eea
The $B \to \pi$ processes follow the same expression with proper replacement of particle mass, lifetime, CKM matrix elements and Wilson coefficients. 
%The  $D^0 \to \pi^0$ form factors are scaled as $f_i \to f_i/\sqrt{2}$ by isospin symmetry, where $f_i$ are $D^+ \to \pi^+$ form factors. 
 To compute the branching ratios of $B^{+(0)} \to (\pi^{+(0)},K^{+(0)}) ll$  in the SM, all the required input parameters are taken from \cite{Tanabashi:2018oca}. The  form factors for $B \to (\pi,K)$ in the light cone sum rule approach are considered from \cite{Colangelo:1996ay, Ball:2004ye}.

%=================================================
\subsection{$B \to (\pi, K) \nu_l \bar \nu_l$}
%==================================================
The differential branching ratio of $B \to P \nu_l \bar \nu_l$ process, where, $P=\pi, K$ are pseudoscalar mesons, is given by \cite{Altmannshofer:2009ma} 
\bea
\frac{d{\rm Br}}{ds_B} = \tau_B \frac{G_F^2\alpha_{\rm em}^2}{256\pi^5} | V_{tb}V_{tq}^*|^2 M_B^5 \lambda^{3/2}(s_B,\tilde{M}_P^2,1) |f_+^P(s_B)|^2 \Big|C_L^\nu +C_L^{\nu \rm NP}\Big |^2\,,
\eea
where $\tilde{M}_P = M_P/M_B$ and $s_B = s/M_B^2$.

%==================================================== 
\subsection{$B_{d(s)} \to (K^*,\phi, \rho) \nu_l \bar \nu_l$}
%=====================================================

The double differential decay rate of $B_{d(s)} \to V \nu_l \bar \nu_l$ processes, where, $V=K^*,\phi, \rho$ are the vector mesons, is given by \cite{Altmannshofer:2009ma}
\bea
\frac{d^2 \Gamma}{ds_B d \cos \theta} = \frac{3}{4} \frac{d \Gamma_T}{ds_B } \sin^2 \theta + \frac{3}{2} \frac{d \Gamma_L}{ds_B } \cos^2 \theta\,.
\eea
Here $\Gamma_{L~(T)}$ are  the longitudinal (transverse) part of decay rate 
\bea
\frac{d \Gamma_L}{ds_B } = 3 M_B^2 |A_0|^2,  \hspace{2cm} 
\frac{d \Gamma_T}{ds_B } = 3 M_B^2 (|A_\perp|^2  + |A_\parallel|^2)\,,
\eea
where, the explicit expression for  transversality amplitudes are given as 
\bea
A_\perp(s_B) &=& 2N^{\nu}\sqrt{2} \lambda^{1/2} (1,\tilde{M}_{V}^2, s_B) (C_L^\nu +C_L^{\nu \rm NP}) \frac{V(s_B)}{(1+\tilde{M}_{V})}, \nn \\
A_\parallel (s_B)  &=& -2N^{\nu}\sqrt{2} (1+\tilde{M}_{V}) (C_L^\nu +C_L^{\nu \rm NP} ) A_1 (s_B), \nn \\
A_0 (s_B) &=& -\frac{N^{\nu} (C_L^\nu +C_L^{\nu \rm NP} )}{\tilde{M}_{V}\sqrt{s_B} } \Big [ (1-\tilde{M}_{V}^2 - s_B) (1 + \tilde{M}_{V}) A_1 (s_B)
\nn \\&& \hspace{4cm} - \lambda(1,\tilde{M}_{V}^2, s_B) \frac{A_2 (s_B)}{1 + \tilde{M}_{V}} \Big ],
\eea
with 
\bea
N^{\nu} = V_{tb} V_{tq}^* \left[ \frac{G_F^2 \alpha_{\rm em}^2 M_B^3}{3 \cdot 2^{10} \pi^5 } s_B \lambda^{1/2} (1,\tilde{M}_{V}^2, s_B) \right]^{1/2}, ~~~~~~\tilde{M}_V = M_V/M_B\,.
\eea
For branching ratio computation in the SM,  the  $B_{(s)} \to V$ form factors are taken from \cite{Ball:2004rg} and remaining required input values from PDG \cite{Tanabashi:2018oca}.
%==========================================
\subsection{$\tau \to \mu \nu_\tau \bar \nu_\mu$}
\begin{figure}[htb]
\centering
\includegraphics[width=0.4\linewidth]{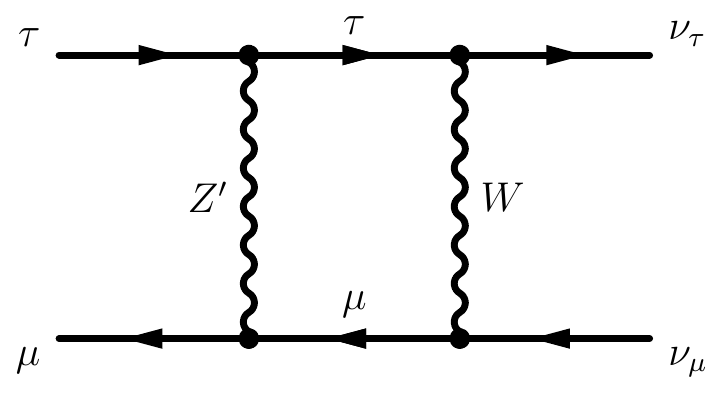}
\caption{Box diagram of $\tau \to \mu \nu_\tau \bar \nu_\mu$ process mediated by the $Z^\prime$ gauge boson.}\label{Fig:taumunu}
\end{figure}

%=================================================
The $\tau \to \mu \nu_\tau \bar \nu_\mu$ process  occur via one loop box diagram in  the presence of $Z^\prime$ boson as shown in Fig. \ref{Fig:taumunu}\,. 

\begin{figure}[htb]
\centering
\includegraphics[width=0.48\linewidth]{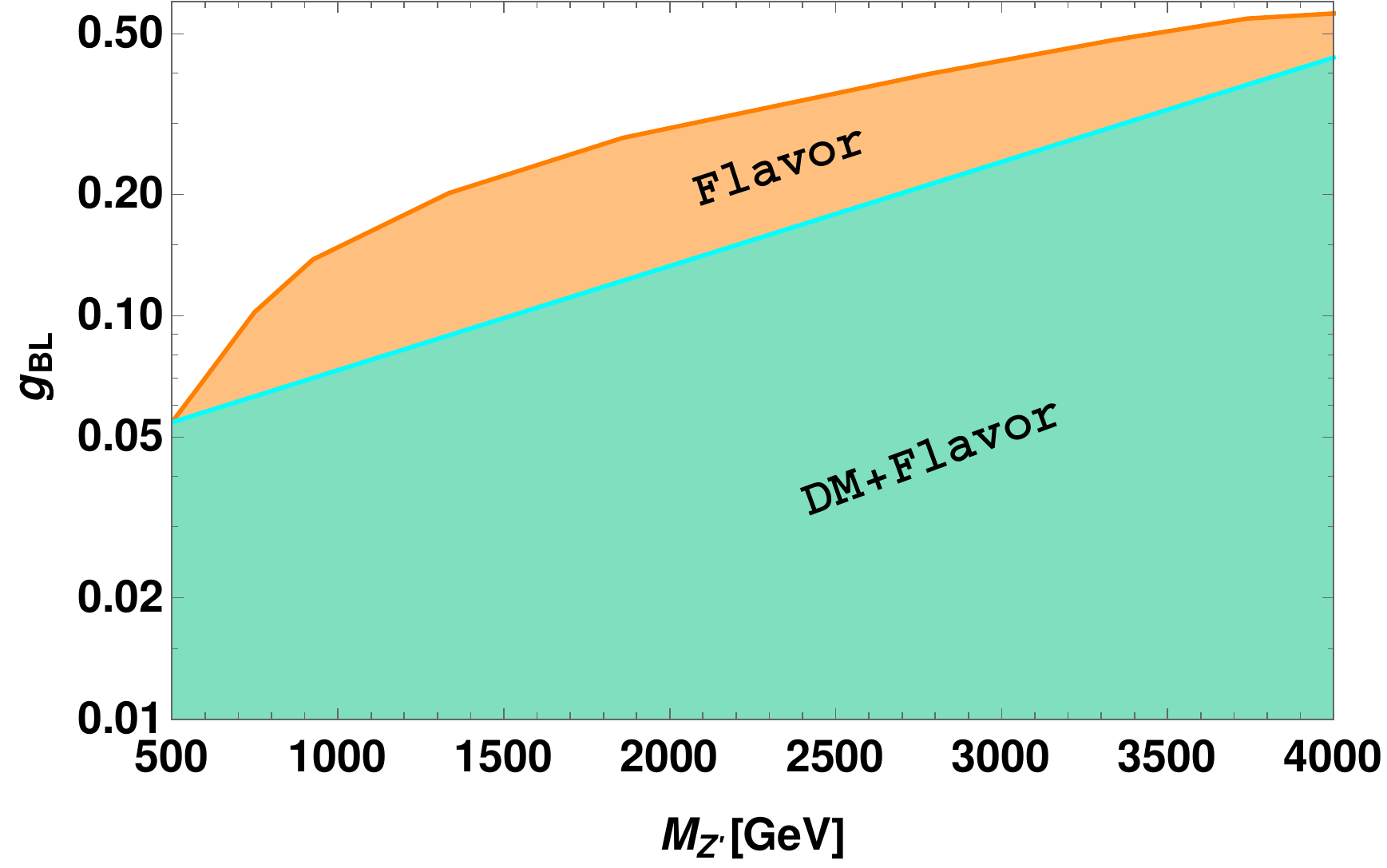}
\caption{Constraints on $M_{Z^\prime}$ and $g_{\rm BL}$ parameters obtained from  DM and flavor observables. Here the orange region is obtained from branching ratios of $b \to q(=s,d) ll(\nu_l \bar \nu_l)$, $\tau \to \mu \nu_\tau \bar \nu_\mu$ processes and the relation $M_{Z^\prime}/g_{\rm BL}>6.9$ TeV. The cyan color represents the region allowed by both DM and flavor studies.}\label{Fig:con-flavor}
\end{figure}

 Including the $Z^\prime$ contribution, the total branching ratio of this process is given by \cite{Altmannshofer:2014cfa}
\bea
{\rm Br}(\tau \to  \mu \nu_\tau \bar \nu_\mu) ={\rm Br}(\tau \to  \mu \nu_\tau \bar \nu_\mu)\big |^{\rm SM} \Bigg ( 1- \frac{3 { g^2_{\rm BL}}}{4\pi^2} \frac{\log (M_W^2/M_{Z^\prime}^2)}{1-M_{Z^\prime}^2/M_W^2} \Bigg )\,.
\eea
 The SM values and corresponding measurements of all the above defined  processes involved in our analysis are presented in Table \ref{Tab:Exp-Flavor}\,.
%===============================================
\begin{table}[htb]
\begin{center}
\begin{tabular}{|c|c|c|}
\hline
~Decay modes~&~SM Values~&~Experimental Limit \cite{Tanabashi:2018oca}~\\
\hline
\hline
~$B_d^0 \to \pi^0 \nu_l \bar \nu_l$~&~$  (3.252\pm 0.26) \times 10^{-6}$~&~$<9\times 10^{-6}$~\\
~$B_d^0 \to K^0 \nu_l \bar \nu_l$~&~$ (4.55\pm 0.342) \times 10^{-6}$~&~$<2.6\times 10^{-5}$~\\
~$B_d^0 \to K^{*0} \nu_l \bar \nu_l$~&~$(9.54\pm 0.66) \times 10^{-6}$~&~$<1.8\times 10^{-5}$~\\
~$B_d^0 \to \phi^0 \nu_l \bar \nu_l$~&~$ (1.11\pm 0.089) \times 10^{-5}$~&~$<1.27\times 10^{-4}$~\\

~$B_d^0 \to \rho^0 \nu  \bar \nu$ &$ (7.624\pm 0.587) \times 10^{-6}$ & $\textless 4.0 \times 10^{-5}$\\
\hline
~$B_u^+ \to \pi^+ \nu_l \bar \nu_l$~&~$ (1.25\pm 0.099) \times 10^{-7}$~&~$<1.4\times 10^{-5}$~\\

~$B_u^+ \to K^+ \nu_l \bar \nu_l$~&~$ (1.752\pm 0.128) \times 10^{-7}$~&~$<1.6\times 10^{-5}$~\\
~$B_u^+ \to K^{* +} \nu  \bar \nu$ & $ (1.03\pm 0.08) \times 10^{-5}$ & $\textless 4.0 \times 10^{-5}$\\

~$B_u^+ \to \rho^+ \nu  \bar \nu$ &$ (8.16\pm 0.653) \times 10^{-6}$ & $\textless 3.0 \times 10^{-5}$\\

\hline
~$B^0 \to \pi^0 e^+ e^-$~&~$(7.66\pm 0.62)\times 10^{-10}$~&~$<8.4\times 10^{-8}$\\
~$B_d^0 \to \pi^0 \mu^+ \mu^-$~&~$(7.67\pm 0.575)\times 10^{-10}$~&~$6.9\times 10^{-8}$\\
~$B_d^0 \to K^0 e^+ e^-$~&~$(1.53\pm 0.1224)\times 10^{-7}$~&~$(1.6^{+1.0}_{-0.8})\times 10^{-7}$\\
~$B_d^0 \to K^0 \mu^+ \mu^-$~&~$(1.51\pm 0.1163)\times 10^{-7}$~&~$(3.39\pm 0.34)\times 10^{-7}$~\\

\hline
~$B_u^+ \to \pi^+ e^+ e^-$~&~$(8.26\pm 0.645)\times 10^{-10}$~&~$<8.0\times 10^{-8}$~\\

~$B_u^+ \to \pi^+ \mu^+ \mu^-$~&~$(8.27\pm 0.579)\times 10^{-10}$~&~$(1.76\pm 0.23)\times 10^{-8}$~\\

~$B_u^+ \to K^+ e^+ e^-$~&~$(1.643\pm 0.127)\times 10^{-7}$~&~$(5.5\pm 0.7)\times 10^{-7}$~\\

~$B_u^+ \to K^+ \mu^+ \mu^-$~&~$(1.626\pm 0.122)\times 10^{-7}$~&~$(4.41\pm 0.23)\times 10^{-7}$~\\

~$B_u^+ \to K^+ \tau^+ \tau^-$~&~$(1.54\pm 0.13)\times 10^{-7}$~&~$<2.25\times 10^{-3}$~~\\
\hline
~$\tau \to \mu \nu_\tau \bar \nu_\mu$~&~$(17.29\pm 0.032)\%$ \cite{Altmannshofer:2014cfa}~&~$(17.39\pm 0.04)\%$~\\
\hline
\end{tabular}
\caption{The SM values and the respective experimental limits on the branching ratios of rare $B$ and $\tau$ decay modes. }\label{Tab:Exp-Flavor}
\end{center}
\end{table}
%===================================================
 The exchange of $Z^\prime$ boson  provides only $C_9^{ \rm NP}$ additional contributions to $b \to (s,d) ll$, thus the leptonic    $B_{d,s} \to ll$ decays   could not provide any strict bound on the new parameters. Since the considered model  has no   $Z^\prime l_i l_j$ couplings,  the neutral and charged lepton flavor violating decay processes like $B \to K^{(*)} l_i^\mp l_i^\pm$, $l_i \to l_j \gamma$, $l_i \to l_j l_k \bar l_k$ do not play any role.   Now using the existing limits on the branching ratios of allowed decay modes (Table \ref{Tab:Exp-Flavor}) and applying the relation $M_{Z^\prime}/g_{\rm BL}>6.9 ~{\rm TeV}$, the constraints on $g_{\rm BL}$ and $M_{Z^\prime}$ parameters are shown in orange color in Fig. \ref{Fig:con-flavor} (Flavor). In this figure, the parameter space allowed by both DM and flavor studies (DM+Flavor) are graphically presented in cyan color. From Fig. \ref{Fig:con-flavor}\,, the bound on $\frac{M_{Z^\prime}}{g_{\rm BL}}$ is found to be greater than $7.14~ (9.1)$ TeV from Flavor (DM+Flavor) case.

%\begin{table}[htb]
%\begin{center}
%\begin{tabular}{|c|c|c|}
%\hline
%~Parameters~&~Flavor~&~DM+Flavor~\\
%\hline\hline
%~$M_{Z^\prime}$ [GeV]~&~$500-4000$ ~&~$500-4000$ ~\\
%~$ g_{\rm BL}$~&~$0-0.56$~&~$0-0.435$~\\
%\hline
%\end{tabular}
%\caption{Predicted allowed ranges of $M_{Z^\prime}$ and $ g_{\rm BL}$ parameters obtained from only flavor observables and  both from dark matter and flavor studies. }\label{Tab:con}
%\end{center}
%\end{table}
\section{Conclusion}

In this article, we have addressed dark matter phenomenology, leptogenesis, light neutrino mass and rare $B$ decay modes in a simple $\rm U(1)_{B-L}$ gauge extension of standard model. Four exotic fermions with fractional $\rm B-L$ charges are included to make the model free of triangle gauge anomalies. We have computed relic density and direct detection cross section of the singlet scalar, whose stability is assured by the $Z_2$ symmetry. The channels contributing to relic density are mediated by scalars and $Z^\prime$ boson. We have constrained the new parameters of the proposed model, by imposing  
Planck Satellite data  on relic density and PandaX limit  on spin independent DM-nucleon scattering cross-section. Along with, we obtained  the constraints on the $Z^\prime$ mass and the gauge coupling from LEP-II and ATLAS dilepton study. 
    
This interesting model can accommodate the explanation for lepton asymmetry with a dimension five Dirac interaction. We considered the resonant enhancement in CP asymmetry with quasi degenerate heavy fermions. We have obtained the lepton asymmetry by solving the Boltzmann equations governing the particle dynamics, which is compatible with the observed baryon asymmetry  $\approx \mathcal{O} (10^{-10})$. Further, we have discussed the flavor effects by computing the asymmetries in each lepton flavor sector. Neutrino mass is realized using type-I seesaw. We found that Yukawa ($\tilde{Y^\prime_{ij}}$) to be of order $10^{-7}$ gives a consistent picture in the perspective of oscillation data and observed baryon asymmetry of the Universe.

We have imposed additional constraint on the new gauge parameters from the available data in the quark and lepton sectors. Since there is no new contribution to $C_{10}$ coefficient, one could not constrain the new parameters from the leptonic $B$ decay modes.  The $Z^\prime$ boson has no lepton flavor violating couplings, thus  the channels like $B \to K^{(*)} l_i l_j$, $l_i \to l_j \gamma$ and $l_i \to 3 l_j$ do not play any role. Hence, we have only considered the branching ratios of rare semileptonic lepton flavor conserving $B$ and $\tau$ decays to compute the allowed parameter space. To conclude, the proposed model provides a common platform to address various phenomenological aspects compatible with their respective current experimental bounds.

 %    \blue{ Similar $\rm B-L$ charges of DM and the extra singlet scalar helps in generating a mass splitting between the CP odd and even component of DM scalar. Unlike the case of scalar triplet extension which explain the radiative generation of neutrino mass due to the effective Dirac interaction with the DM, the scalar singlet extension can explain a tree level neutrino mass without any common parameter with the DM sector. Hence we realize the importance of two different extension of the four exotic fermion model, which involves the cases of scalar triplet and a scalar singlet. In the former case neutrino mass and DM has a common parameter space with radiative generation of neutrino mass, while the later involves a common parameter space for neutrino mass and leptogenesis and DM doesn't have any common coupling with these two sectors. }

\acknowledgments

SM would like to thank DST Inspire for the financial support. We acknowledge Prof. Anjan Giri, Prof. Rukmani Mohanta, Dr. Narendra Sahu and  Dr. Sudhanwa Patra for their useful discussions towards this work.

\bibliography{BL}

\end{document}